\documentclass[%
reprint,
superscriptaddress,
showpacs,
 amsmath,amssymb,
 aps,
]{revtex4-1}  

\usepackage{graphicx}
\usepackage{dcolumn}
\usepackage{bm}
\usepackage{amsmath}
\usepackage{autobreak}
\usepackage{hyperref}
\usepackage[mathlines]{lineno}
\usepackage{natbib}
\usepackage{multirow}
\usepackage{booktabs}

\begin{document}

\preprint{APS/123-QED}

\title{First search for light fermionic dark matter absorption on electrons using germanium detector in CDEX-10 experiment}

\author{J.~X.~Liu}
\affiliation{Key Laboratory of Particle and Radiation Imaging (Ministry of Education) and Department of Engineering Physics, Tsinghua University, Beijing 100084}
\author{L.~T.~Yang}
\email{Corresponding author: yanglt@mail.tsinghua.edu.cn}
\affiliation{Key Laboratory of Particle and Radiation Imaging (Ministry of Education) and Department of Engineering Physics, Tsinghua University, Beijing 100084}
\author{Q.~Yue}
\email{Corresponding author: yueq@mail.tsinghua.edu.cn}
\affiliation{Key Laboratory of Particle and Radiation Imaging (Ministry of Education) and Department of Engineering Physics, Tsinghua University, Beijing 100084}
\author{K.~J.~Kang}
\affiliation{Key Laboratory of Particle and Radiation Imaging (Ministry of Education) and Department of Engineering Physics, Tsinghua University, Beijing 100084}
\author{Y.~J.~Li}
\affiliation{Key Laboratory of Particle and Radiation Imaging (Ministry of Education) and Department of Engineering Physics, Tsinghua University, Beijing 100084}

\author{H.~P.~An}
\affiliation{Key Laboratory of Particle and Radiation Imaging (Ministry of Education) and Department of Engineering Physics, Tsinghua University, Beijing 100084}
\affiliation{Department of Physics, Tsinghua University, Beijing 100084}

\author{Greeshma~C.}
\altaffiliation{Participating as a member of TEXONO Collaboration}
\affiliation{Institute of Physics, Academia Sinica, Taipei 11529}

\author{J.~P.~Chang}
\affiliation{NUCTECH Company, Beijing 100084}

\author{Y.~H.~Chen}
\affiliation{YaLong River Hydropower Development Company, Chengdu 610051}
\author{J.~P.~Cheng}
\affiliation{Key Laboratory of Particle and Radiation Imaging (Ministry of Education) and Department of Engineering Physics, Tsinghua University, Beijing 100084}
\affiliation{School of Physics and Astronomy, Beijing Normal University, Beijing 100875}
\author{W.~H.~Dai}
\affiliation{Key Laboratory of Particle and Radiation Imaging (Ministry of Education) and Department of Engineering Physics, Tsinghua University, Beijing 100084}
\author{Z.~Deng}
\affiliation{Key Laboratory of Particle and Radiation Imaging (Ministry of Education) and Department of Engineering Physics, Tsinghua University, Beijing 100084}
\author{C.~H.~Fang}
\affiliation{College of Physics, Sichuan University, Chengdu 610065}
\author{X.~P.~Geng}
\affiliation{Key Laboratory of Particle and Radiation Imaging (Ministry of Education) and Department of Engineering Physics, Tsinghua University, Beijing 100084}
\author{H.~Gong}
\affiliation{Key Laboratory of Particle and Radiation Imaging (Ministry of Education) and Department of Engineering Physics, Tsinghua University, Beijing 100084}
\author{Q.~J.~Guo}
\affiliation{School of Physics, Peking University, Beijing 100871}
\author{T.~Guo}
\affiliation{Key Laboratory of Particle and Radiation Imaging (Ministry of Education) and Department of Engineering Physics, Tsinghua University, Beijing 100084}
\author{X.~Y.~Guo}
\affiliation{YaLong River Hydropower Development Company, Chengdu 610051}
\author{L.~He}
\affiliation{NUCTECH Company, Beijing 100084}
\author{J.~R.~He}
\affiliation{YaLong River Hydropower Development Company, Chengdu 610051}
\author{J.~W.~Hu}
\affiliation{Key Laboratory of Particle and Radiation Imaging (Ministry of Education) and Department of Engineering Physics, Tsinghua University, Beijing 100084}
\author{H.~X.~Huang}
\affiliation{Department of Nuclear Physics, China Institute of Atomic Energy, Beijing 102413}
\author{T.~C.~Huang}
\affiliation{Sino-French Institute of Nuclear and Technology, Sun Yat-sen University, Zhuhai 519082}
\author{L.~Jiang}
\affiliation{Key Laboratory of Particle and Radiation Imaging (Ministry of Education) and Department of Engineering Physics, Tsinghua University, Beijing 100084}
\author{S.~Karmakar}
\altaffiliation{Participating as a member of TEXONO Collaboration}
\affiliation{Institute of Physics, Academia Sinica, Taipei 11529}

\author{H.~B.~Li}
\altaffiliation{Participating as a member of TEXONO Collaboration}
\affiliation{Institute of Physics, Academia Sinica, Taipei 11529}
\author{H.~Y.~Li}
\affiliation{College of Physics, Sichuan University, Chengdu 610065}
\author{J.~M.~Li}
\affiliation{Key Laboratory of Particle and Radiation Imaging (Ministry of Education) and Department of Engineering Physics, Tsinghua University, Beijing 100084}
\author{J.~Li}
\affiliation{Key Laboratory of Particle and Radiation Imaging (Ministry of Education) and Department of Engineering Physics, Tsinghua University, Beijing 100084}
\author{M.~C.~Li}
\affiliation{YaLong River Hydropower Development Company, Chengdu 610051}
\author{Q.~Y.~Li}
\affiliation{College of Physics, Sichuan University, Chengdu 610065}
\author{R.~M.~J.~Li}
\affiliation{College of Physics, Sichuan University, Chengdu 610065}
\author{X.~Q.~Li}
\affiliation{School of Physics, Nankai University, Tianjin 300071}
\author{Y.~L.~Li}
\affiliation{Key Laboratory of Particle and Radiation Imaging (Ministry of Education) and Department of Engineering Physics, Tsinghua University, Beijing 100084}
\author{Y.~F.~Liang}
\affiliation{Key Laboratory of Particle and Radiation Imaging (Ministry of Education) and Department of Engineering Physics, Tsinghua University, Beijing 100084}
\author{B.~Liao}
\affiliation{School of Physics and Astronomy, Beijing Normal University, Beijing 100875}
\author{F.~K.~Lin}
\altaffiliation{Participating as a member of TEXONO Collaboration}
\affiliation{Institute of Physics, Academia Sinica, Taipei 11529}
\author{S.~T.~Lin}
\affiliation{College of Physics, Sichuan University, Chengdu 610065}
\author{S.~K.~Liu}
\affiliation{College of Physics, Sichuan University, Chengdu 610065}
\author{Y.~D.~Liu}
\affiliation{School of Physics and Astronomy, Beijing Normal University, Beijing 100875}
\author{Y.~Liu}
\affiliation{College of Physics, Sichuan University, Chengdu 610065}
\author{Y.~Y.~Liu}
\affiliation{School of Physics and Astronomy, Beijing Normal University, Beijing 100875}
\author{H.~Ma}
\affiliation{Key Laboratory of Particle and Radiation Imaging (Ministry of Education) and Department of Engineering Physics, Tsinghua University, Beijing 100084}
\author{Y.~C.~Mao}
\affiliation{School of Physics, Peking University, Beijing 100871}
\author{Q.~Y.~Nie}
\affiliation{Key Laboratory of Particle and Radiation Imaging (Ministry of Education) and Department of Engineering Physics, Tsinghua University, Beijing 100084}
\author{H.~Pan}
\affiliation{NUCTECH Company, Beijing 100084}
\author{N.~C.~Qi}
\affiliation{YaLong River Hydropower Development Company, Chengdu 610051}
\author{J.~Ren}
\affiliation{Department of Nuclear Physics, China Institute of Atomic Energy, Beijing 102413}
\author{X.~C.~Ruan}
\affiliation{Department of Nuclear Physics, China Institute of Atomic Energy, Beijing 102413}
\author{M.~B.~Shen}
\affiliation{YaLong River Hydropower Development Company, Chengdu 610051}
\author{M.~K.~Singh}
\altaffiliation{Participating as a member of TEXONO Collaboration}
\affiliation{Institute of Physics, Academia Sinica, Taipei 11529}
\affiliation{Department of Physics, Banaras Hindu University, Varanasi 221005}
\author{T.~X.~Sun}
\affiliation{School of Physics and Astronomy, Beijing Normal University, Beijing 100875}
\author{W.~L.~Sun}
\affiliation{YaLong River Hydropower Development Company, Chengdu 610051}
\author{C.~J.~Tang}
\affiliation{College of Physics, Sichuan University, Chengdu 610065}
\author{Y.~Tian}
\affiliation{Key Laboratory of Particle and Radiation Imaging (Ministry of Education) and Department of Engineering Physics, Tsinghua University, Beijing 100084}
\author{G.~F.~Wang}
\affiliation{School of Physics and Astronomy, Beijing Normal University, Beijing 100875}
\author{J.~Z.~Wang}
\affiliation{Key Laboratory of Particle and Radiation Imaging (Ministry of Education) and Department of Engineering Physics, Tsinghua University, Beijing 100084}
\author{L.~Wang}
\affiliation{School of Physics and Astronomy, Beijing Normal University, Beijing 100875}
\author{Q.~Wang}
\affiliation{Key Laboratory of Particle and Radiation Imaging (Ministry of Education) and Department of Engineering Physics, Tsinghua University, Beijing 100084}
\affiliation{Department of Physics, Tsinghua University, Beijing 100084}
\author{Y.~F.~Wang}
\affiliation{Key Laboratory of Particle and Radiation Imaging (Ministry of Education) and Department of Engineering Physics, Tsinghua University, Beijing 100084}
\author{Y.~X.~Wang}
\affiliation{School of Physics, Peking University, Beijing 100871}
\author{H.~T.~Wong}
\altaffiliation{Participating as a member of TEXONO Collaboration}
\affiliation{Institute of Physics, Academia Sinica, Taipei 11529}

\author{Y.~C.~Wu}
\affiliation{Key Laboratory of Particle and Radiation Imaging (Ministry of Education) and Department of Engineering Physics, Tsinghua University, Beijing 100084}
\author{H.~Y.~Xing}
\affiliation{College of Physics, Sichuan University, Chengdu 610065}
\author{K.~Z.~Xiong}
\affiliation{YaLong River Hydropower Development Company, Chengdu 610051}
\author{R. Xu}
\affiliation{Key Laboratory of Particle and Radiation Imaging (Ministry of Education) and Department of Engineering Physics, Tsinghua University, Beijing 100084}
\author{Y.~Xu}
\affiliation{School of Physics, Nankai University, Tianjin 300071}
\author{T.~Xue}
\affiliation{Key Laboratory of Particle and Radiation Imaging (Ministry of Education) and Department of Engineering Physics, Tsinghua University, Beijing 100084}
\author{Y.~L.~Yan}
\affiliation{College of Physics, Sichuan University, Chengdu 610065}
\author{N.~Yi}
\affiliation{Key Laboratory of Particle and Radiation Imaging (Ministry of Education) and Department of Engineering Physics, Tsinghua University, Beijing 100084}
\author{C.~X.~Yu}
\affiliation{School of Physics, Nankai University, Tianjin 300071}
\author{H.~J.~Yu}
\affiliation{NUCTECH Company, Beijing 100084}
\author{M.~Zeng}
\affiliation{Key Laboratory of Particle and Radiation Imaging (Ministry of Education) and Department of Engineering Physics, Tsinghua University, Beijing 100084}
\author{Z.~Zeng}
\affiliation{Key Laboratory of Particle and Radiation Imaging (Ministry of Education) and Department of Engineering Physics, Tsinghua University, Beijing 100084}
\author{B.~T.~Zhang}
\affiliation{Key Laboratory of Particle and Radiation Imaging (Ministry of Education) and Department of Engineering Physics, Tsinghua University, Beijing 100084}
\author{F.~S.~Zhang}
\affiliation{School of Physics and Astronomy, Beijing Normal University, Beijing 100875}
\author{L.~Zhang}
\affiliation{College of Physics, Sichuan University, Chengdu 610065}
\author{P.~Zhang}
\affiliation{YaLong River Hydropower Development Company, Chengdu 610051}
\author{Z.~H.~Zhang}
\affiliation{Key Laboratory of Particle and Radiation Imaging (Ministry of Education) and Department of Engineering Physics, Tsinghua University, Beijing 100084}
\author{Z.~Y.~Zhang}
\affiliation{Key Laboratory of Particle and Radiation Imaging (Ministry of Education) and Department of Engineering Physics, Tsinghua University, Beijing 100084}
\author{J.~Z.~Zhao}
\affiliation{Key Laboratory of Particle and Radiation Imaging (Ministry of Education) and Department of Engineering Physics, Tsinghua University, Beijing 100084}
\author{K.~K.~Zhao}
\affiliation{College of Physics, Sichuan University, Chengdu 610065}
\author{M.~G.~Zhao}
\affiliation{School of Physics, Nankai University, Tianjin 300071}

\author{J.~F.~Zhou}
\affiliation{YaLong River Hydropower Development Company, Chengdu 610051}
\author{Z.~Y.~Zhou}
\affiliation{Department of Nuclear Physics, China Institute of Atomic Energy, Beijing 102413}
\author{J.~J.~Zhu}
\affiliation{College of Physics, Sichuan University, Chengdu 610065}

\collaboration{CDEX Collaboration}
\noaffiliation
\date{\today}

\begin{abstract}
We present the first results of the search for sub-MeV fermionic dark matter absorbed by electron targets of germanium using the 205.4~kg$\cdot$day data collected by the CDEX-10 experiment, with the analysis threshold of 160~eVee. No significant dark matter (DM) signals over the background are observed. Results are presented as limits on the cross section of DM--electron interaction. We present new constraints of cross section in the DM range of 0.1--10 keV/$c^2$ for vector and axial-vector interaction. The upper limit on the cross section is set to be $\rm 6.8\times10^{-46}~cm^2$ for vector interaction, and $\rm 2.3\times10^{-46}~cm^2$ for axial-vector interaction at DM mass of 5 keV/$c^2$.
\end{abstract}

\maketitle

\emph{Introduction.}— 
Various astronomical and cosmological observations support the existence of dark matter (DM), which accounts for approximately 26.8\% of the Universe~\cite{DM_Universe}. As the most popular candidates for DM, weakly interacting massive particles (WIMPs) have been searched by a lot of direct detection experiments, such as XENON~\cite{XENONnT_2023}, LUX-ZEPLIN~\cite{LZ_2023}, PandaX~\cite{PandaX-4T_2023}, DarkSide~\cite{DarkSide}, SuperCDMS~\cite{SuperCDMS}, and CDEX~\cite{CDEX_1_LSK,CDEX_2_ZW,CDEX_3_YQ,CDEX_4_ZW,CDEX_5_YLT,CDEX_6_JH,CDEX_7_JH,CDEX_8_YLT,CDEX_9_LZZ}. These direct detection experiments have set stringent limits on the cross section of DM interaction with Standard Model (SM) particles within the DM mass range of GeV/$c^2$ to TeV/$c^2$ via DM--nucleus scattering ($\chi$--$N$). Meanwhile, the cross section of sub-GeV DM interaction with SM particles is less constrained. Recently, physics channels, such as the Migdal effect of $\chi$--$N$~\cite{CDEX_9_LZZ,Migdal_2_LUX,Migdal_4_LZZ,Migdal_3_CRESST}, fermionic DM absorbed by nuclear targets $(\chi+A\rightarrow\nu+A)$, DM--nucleus $\rm 3\rightarrow2$ scattering $(\chi+\chi+A\rightarrow\phi+A)$~\cite{FDNA_1_Dror,FDNA_2_Dror,FDNA_3_PandaX,FDNA_4_MJD,FDNA_5_DWH,3to2_CW,FDMD_Li}, and DM--electron scattering ($\chi$--$e$)~\cite{chi-e_DAMIC,chi-e_DarkSide,chi-e_EDELWEISS,chi-e_PandaX,chi-e_SENSEI,chi-e_SuperCDMS,chi-e_XENON,chi-e_XENON_2,CDEX_ZZY}, have been explored to search for DM in the MeV/$c^2$ mass scale [$m_\chi\sim\mathcal{O}$(MeV/$c^2$)]. 

However, if we want to explore lighter (sub-MeV) DM, particularly DM of few keV/$c^2$ in a direct detection experiment with a typical $\mathcal{O}$(1) keV experimental threshold, considering the aforementioned physics scenarios, the energy deposited by DM in the detector will decrease to below the threshold. There are two strategies to probe lighter DM: (1) increasing the energy deposition from DM in targets and (2) using lower experimental threshold detectors. To increase the energy deposited by DM in targets, DM particles with higher kinetic energy boosted by interaction with SM particles; for example, DM boosted by cosmic rays have been searched~\cite{CRDM_25,CRDM_26,CRDM_27,CRDM_28,CRDM_29,CRDM_30,CRDM_31,CDEX_XR}. On the other hand, considering a new DM interaction paradigm, where the energy transfer from DM particle to SM particle is not dominated by kinetic energy but by rest energy $E=mc^2$, the energy deposit by sub-MeV DM can also overcome the experimental threshold. This is the main motivation to study DM absorption for bosonic DM~\cite{Bosonic_1,Bosonic_2,Bosonic_3,Bosonic_4,Bosonic_5,Bosonic_6,Bosonic_7,Bosonic_8,Bosonic_9,Bosonic_10} (e.g., dark photon and axionlike particle~\cite{Bosonic_9}), and fermionic DM~\cite{FDNA_1_Dror,FDNA_2_Dror,FDEA_2_Dror}. Recently, a new type of interaction where fermionic DM is absorbed by electron targets $(\chi+e^-\rightarrow\nu+e^-)$ has been proposed and searched~\cite{FDEA_1_GSF,FDEA_2_Dror,FDEA_3_PandaX}. When fermionic DM is absorbed by electron target, the electron obtains a kinetic recoil through the mass of $\chi$. As mentioned above, the lower experimental threshold detectors, the lighter DM can be probed.

The CDEX-10 experiment is the second phase of the CDEX experiment operated in the China Jinping Underground Laboratory(CJPL) with a 2400~m rock overburden~\cite{CDEX_7_JH,CDEX_CJP,CDEX_6_JH}. CDEX-10 has three detector arrays named C10A(B, C). Each detector array comprises three p-type point contact germanium (PPCGe) detectors: Ge1, Ge2, and Ge3. The detectors are shielded with 20 cm high-purity oxygen-free copper and immersed in a liquid nitrogen cryostat. Details of the experimental setup can be found in Refs.~\cite{CDEX_7_JH,CDEX_6_JH}. One of the detectors, C10B-Ge1, achieved the lowest analysis threshold (160~eVee, ``eVee''represents the electron equivalent energy derived from energy calibration), with an exposure of 205.4 kg$\cdot$day~\cite{CDEX_SZ}. The data processing procedure includes data quality check, physics event selection, energy calibration, and bulk or surface event discrimination~\cite{CDEX_7_JH,CDEX_BS_YLT}. The 0.16--2.16 keV spectrum after all event selections is used in this analysis. In this paper, we report the search results of keV/$c^2$ scale DM absorbed by electron using 205.4 kg$\cdot$day exposure data from the CDEX-10 experiment with an energy threshold of 160~eVee.

\emph{Expected signal.}— 
Fermionic DM is absorbed by the electron target via the following process:
\begin{equation}
    \chi + e^- \rightarrow \nu + e^-,
\end{equation}
where a DM particle $\chi$ is absorbed by an electron and a neutrino is emitted~\cite{FDEA_1_GSF,FDEA_2_Dror}.

In this analysis, for effective interaction, the vector type and axial-vector type operators are considered and can be written as follows~\cite{FDEA_1_GSF}:
\begin{equation}
    \begin{aligned}
        \mathcal{O}^{V}_{e\mu \chi} = \frac{1}{\Lambda^2}(\overline{e}\gamma_\mu e)(\overline{\nu}_L\gamma^\mu\chi_L),\\
        \mathcal{O}^{A}_{e\mu \chi} = \frac{1}{\Lambda^2}(\overline{e}\gamma_\mu\gamma_5 e)(\overline{\nu}_L\gamma^\mu\chi_L),    
    \end{aligned}
\end{equation}
where the DM is treated as a Dirac fermion with lepton number 1~\cite{FDNA_1_Dror}, and the SM left-handed neutrino is considered. $1/\Lambda^2$ is the Wilson coefficient, with a dimension of $\rm [mass]^{-2}$.  

Considering the standard halo model, the DM velocity follows the Maxwell-Boltzmann distribution with a cutoff at the Galactic escape velocity $v_{\chi max}\sim 544~\rm km/s$, indicating that the DM is nonrelativistic. In the DM absorption process, the entire DM mass is released as energy of the final-state particles. The kinetic energy of DM $T_\chi\approx\frac{1}{2}m_\chi v_\chi^2$ can be neglected, $T_\chi/m_\chi\sim\mathcal{O}(10^{-6})$~\cite{FDEA_1_GSF}. The law of energy conservation gives
\begin{equation}
    m_\chi+E^{nl}_B = q+E_R,
\end{equation}
where $m_\chi$ is the mass of DM $\chi$, $E^{nl}_B ~(E^{nl}_B < 0)$ is the binding energy of the electron in the initial state $(n,l)$ shell, $q$ is the neutrino energy, and $E_R$ is the electron recoil energy in the final state.

The total detectable energy $E_{det}$ can be expressed as follows:
\begin{equation}
    E_{det} = E_R + |E^{nl}_B|,
\end{equation}
where $E_R$ is the electron recoil energy and $|E^{nl}_B|$ is the deexcitation energy. The deexcitation process contains characteristic x-rays and Auger electrons, with a total energy of $|E^{nl}_B|$.

Following the procedures in Refs.~\cite{FDEA_1_GSF,FDEA_2_Dror}, the expected differential event rate of the detectable energy can be expressed as follows: 
\begin{equation}
    \frac{dR}{dE_{det}} = N_T\frac{\rho_\chi}{m_\chi}\sum\limits_{n,l}{\frac{|\mathcal{M}(q)|^2}{64\pi m_\chi m_e^2}\frac{q}{E_{det} - |E^{nl}_B|}|f^{nl}_{ion}(k^{'},q)|^2},
\end{equation}
where $N_T$ is the number of Ge atoms, $\rho_\chi$ is the local DM density taken as 0.3~GeV/$c^2$/cm$^3$ ~\cite{DM_rho_1,DM_rho_2}, $m_e$ is the electron mass, and $|f^{nl}_{ion}(k^{'},q)|^2$ is the ionization form factor for an atomic orbit $(n,l)$ ~\cite{fion_1,FDEA_2_Dror}, which is calculated by modifying the public code $\tt DARKART$~\cite{fion_1,DarkART_1}. $k^{'}=\sqrt{2m_e \cdot (E_{det} - |E^{nl}_B|)}$ is the final momentum of the outgoing electron, and $|\mathcal{M}(q)|^2$ is the scattering matrix element, which can be expressed as follows:
\begin{equation}
    |\mathcal{M}^{V,A}(q)|^2 = (4, 12)\times\frac{4\pi m_e^2 q}{m_\chi}(\sigma_e v_\chi),
\end{equation}
for $\mathcal{O}^{V}_{e\mu \chi}$ and $\mathcal{O}^{A}_{e\mu \chi}$, respectively~\cite{FDEA_1_GSF}. $\sigma_e$ is the total scattering cross section between the DM and the free electron. For the tiny mass DM ($m_\chi \ll m_e$), $\sigma_e$ can be expressed as follows:
\begin{equation}
    \sigma_e = \frac{m_\chi^2}{4\pi\Lambda^4 v_\chi},
\end{equation}

\begin{figure}[!htbp] 
	\includegraphics[width=\linewidth]{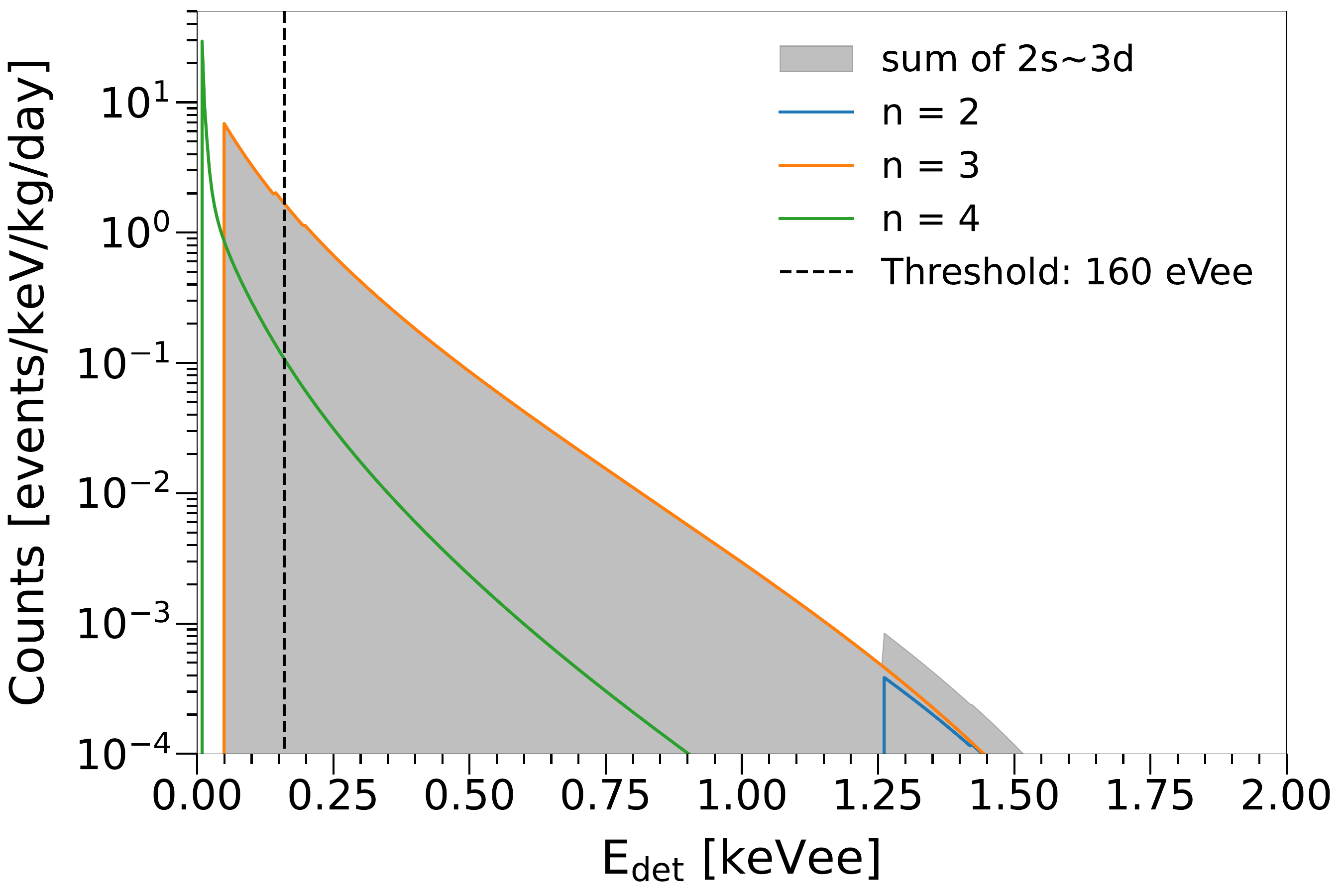}
\caption{Predicted visible energy spectra of the fermionic DM absorbed by electron targets of Ge for a vector-type operator for $m_\chi$ = 2 keV/$c^2$ and $\sigma_e v_\chi$ = $10^{-45}$~cm$^2$. Contributions by different shells of electrons are shown in different colors. The signals used in this analysis (summation from $1s$ to $3d$) are shown in gray shade. In this scenario with $m_\chi$ = 2 keV/$c^2$, $K$-shell ($n$=1) electrons cannot be ionized due to the small $m_\chi$. The $N$-shell ($n$=4) spectra shown here is calculated assuming Ge atoms as isolated atoms, but the contribution from the $N$ shell is not included. The energy resolution is not considered in this figure.} 
\label{fig_theo_spec_vecotr} 
\end{figure}

Considering Ge, the expected energy spectra of fermionic DM absorbed by the electron target via a vector interaction are shown in Fig.~\ref{fig_theo_spec_vecotr}, with $m_\chi$ = 2 keV/$c^2$ and $\sigma_e v_\chi$ = $10^{-45}$~cm$^2$. 

Notably, the electronic band structures of Ge crystal are not considered in this study. Ge atoms are treated as isolated atoms. Only the contribution from core electrons, i.e., $K$, $L$, and $M$ shells $(n = 1,2,3)$, are considered. Because the electronic band structures near the band gap ($N$ shell, $n=4$, valence and conduction electronic states as stated in Ref.~\cite{crystal_1_Griffin}) in Ge crystal are non-negligible and do not fit the assumption of isolated atoms. Treating Ge atoms as isolated atoms and only the contribution from core electrons ($K$, $L$, and $M$ shells) are considered make our results conservative.

\begin{figure}[!htbp] 
\includegraphics[width=\linewidth]{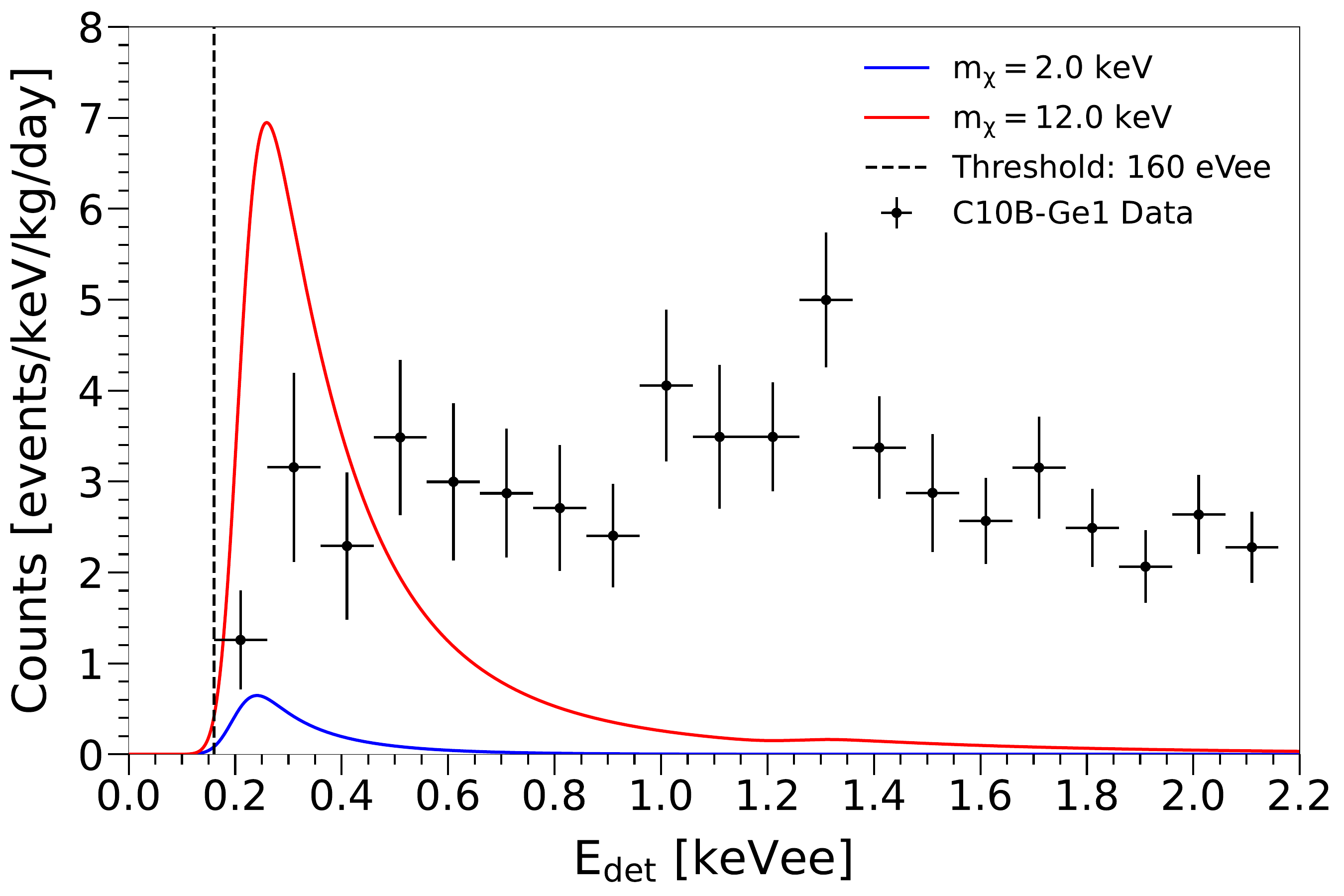}
\caption{Expected HPGe detector response in the vector interaction case at $\sigma_e v_\chi =\rm 10^{-45}~cm^2$ for $m_\chi =\rm 2~keV/c^2$ (blue) and $m_\chi =\rm 12~keV/c^2$ (red) are superimposed. Experimental data from CDEX-10 before combined efficiency correction in 0.16-2.16 keVee are also depicted. The bin width is 100 eVee.
}
\label{fig_blur_spec}
\end{figure}

In Fig.~\ref{fig_theo_spec_vecotr}, $K$-shell electrons cannot be ionized due to small $m_\chi$. When $m_\chi > |E_B^{1,0}|$, contribution by $K$-shell electrons will be taken into account. The binding energies of electrons in different Ge atom shells are summarized in Table~\ref{table1}. Figure~\ref{fig_blur_spec} shows the spectra used in this analysis considering the detector energy resolution. The energy resolution of C10B-Ge1 is $\sigma(E) = 35.8\times10^{-3} + 16.6\times10^{-3}\sqrt{E}$~\cite{CDEX_XR}, where $E$ and $\sigma(E)$ are in keV.

\begin{table}[!tbp]
\caption{\label{table1}
Binding energies $(E_B^{n,l})$ of the electrons in Ge atom shells $(n, l)$. (Data from Ref.~\cite{bind_eng_Ge})} 
\begin{ruledtabular}
\begin{tabular*}{\linewidth}{cccc}
\multicolumn{4}{c}{$E_B^{n,l} ~ \rm [eV]$}\\
\hline
 & \multicolumn{3}{c}{$l$}\\
 \cline{2-4}
$n$ & $s$ & $p$ & $d$ \\
\hline
1 & $-1.1\times10^{4}$ & ... & ... \\
2 & $-1.4\times10^{3}$ & $-1.3\times10^{3}$ & ... \\
3 & $-2.0\times10^{2}$ & $-1.4\times10^{2}$ & $-44$ \\
4 & $-15$ & $-7.8$ & ... \\
\end{tabular*}
\end{ruledtabular}
\end{table}

\emph{Data analysis.}— 
Data used in this analysis are obtained from CDEX-10, specifically from the C10B-Ge1 detector between February 2017 and August 2018, with an exposure of 205.4 kg$\cdot$day~\cite{CDEX_SZ}. The sensitive mass of C10B-Ge1 is 939 g with a dead layer thickness of 0.88$\pm$0.12 mm~\cite{CDEX_5_YLT,CDEX_7_JH}. After a series of data processing procedures, we derive the physics spectrum. The analysis threshold is set to be 160 eVee, the combined efficiency at 160 eVee is 4.5\%, for the first analysis bin (160--260 eVee), the averaged combined efficiency is 48.0\%. The combined efficiency is composed of 2 parts: physics-noise (PN) cut efficiency and trigger efficiency~\cite{CDEX_6_JH}.

In the interest energy range of 0.16--2.16 keVee, background events are composed of the radioactivity from cosmogenic isotopes inside the detector and Compton scattering events from high-energy gamma rays. The radioactivity from cosmogenic isotopes are $L$- and $M$-shell x-ray peaks of cosmogenic isotopes, including $^{68}$Ge, $^{68}$Ga, $^{65}$Zn, $^{55}$Fe, $^{54}$Mn, and $^{49}$V. The $K$-shell peaks in 4--12 keVee spectrum are fitted to limit the intensity of their corresponding $L$- and $M$- shell peaks using the known $K$/$L$ and $K$/$M$ ratios.~\cite{CDEX_XR,CDEX_6_JH,CDEX_SZ}.

The $\chi^2$ analysis~\cite{chisquare} is applied in this work. The $\chi^2$ function is defined as follows:
\begin{equation}
\begin{aligned}
&\chi^2=\\
&\sum^{N}_{i=1}\frac{\{n_i-\int^{E_i+\Delta E}_{E_i - \Delta E} \epsilon(E; \vec{v})[S(E; m_\chi, \sigma_ev_\chi)+B(E; \vec{u})]dE\}}{\sigma_i^{2}}^2 \\
&+\vec{v}^\intercal \mathbf{Q}^{-1}_v\vec{v} + \vec{u}^\intercal \mathbf{Q}^{-1}_u\vec{u}, 
\end{aligned}
\end{equation}
where $n_i$ is the experimental data in the $i$th energy bin, $S(E; m_\chi, \sigma_ev_\chi)$ is DM signals at energy $E$, $B(E; \vec{u})$ is the expected flat background contributed by the Compton scattering of high-energy $\gamma$ rays and $L$- and $M$-shell x-ray peaks of cosmogenic isotopes, i.e., 
\begin{equation}
\begin{aligned}
    &B(E; \vec{u})=C_0+\sum^{N_L}_{L=1}I_L(u_L)\mathcal{G}(E; E_L, \sigma_L)+\\
    &\sum^{N_M}_{M=1}I_M(u_M)\mathcal{G}(E; E_M, \sigma_M)
\end{aligned}
\end{equation}
where $\mathcal{G}$ is the Gaussian function. $\epsilon(E; \vec{v})$ is the combined efficiency at energy $E$.
\begin{equation}
\begin{aligned}
    \epsilon(E; \vec{v}) = &\frac{1}{4}\{[1+\text{Erf}(\frac{E-v_1}{v_2})] \\ &\times [1+\text{Erf}(\frac{E-v_3}{v_4})]\} 
\end{aligned}
\end{equation}
where $\text{Erf}$ is the error function.
$\sigma_i$ is the uncorrelated uncertainties in the $i$th energy bin. $\vec{v}$ and $\vec{u}$ are the nuisance parameters for combined efficiency and radioactive isotope intensities, separately. The nuisance parameters $\vec{v}$ and $\vec{u}$ are summarized in Table~\ref{table2}. $\mathbf{Q}$ is the covariance matrix of nuisance parameters.

\begin{table}[!tbp]
\caption{\label{table2}
Summary of nominal values, uncertainties (fractional) for the radioactive isotopes intensities and combined efficiency parameters.} 
\begin{ruledtabular}
\begin{tabular*}{\linewidth}{cccc}
\multicolumn{4}{c}{Radioactive isotopes intensities}\\
\multicolumn{4}{c}{Unit: events/keV/kg/day}\\

Name & Related symbol & Center of ${I_K}$ & $\sigma_{I_K}$\\
\hline
$^{65}$Zn & $u_1$ & 14.15 & 8.5\% \\
$^{68}$Ga & $u_2$ & 2.12 & 36\% \\
$^{68}$Ge & $u_3$ & 25.94 & 6.6\% \\
$^{55}$Fe & $u_4$ & 6.06 & 16\% \\
$^{54}$Mn & $u_5$ & 2.07 & 36\% \\
$^{49}$V & $u_6$ & 1.72 & 42\% \\

\hline
\hline

\multicolumn{4}{c}{Efficiency parameters }\\

Name & Related symbol & Center of ${v}$ & $\sigma_{v}$\\
\hline
trigger center & $v_1$ & 0.18 & 1.5\% \\
trigger slope & $v_2$ & 0.053 & 18\% \\
PN Cut center & $v_3$ & 0.20 & 0.95\% \\
PN Cut slope & $v_4$ & 0.054 & 5.4\% \\

\end{tabular*}
\end{ruledtabular}
\end{table}

\begin{figure}[!htbp]
    \centering
    \includegraphics[width=\linewidth]{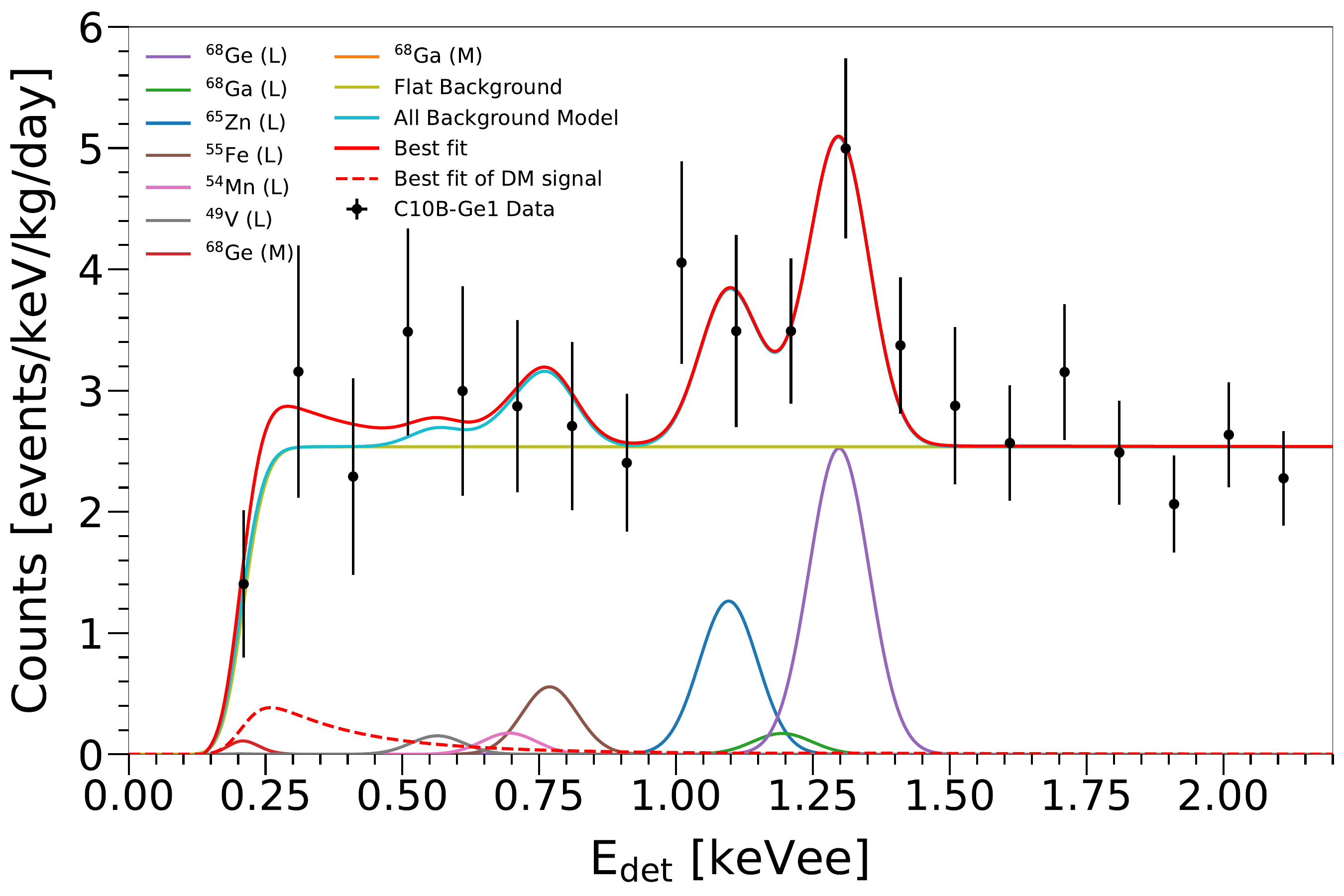}
    \caption{The spectrum from CDEX-10 (black point with error bar), together with the best-fit signal (red dotted line) for $m_\chi$ = 12 keV/$c^2$. The light blue solid line is the best fit for the background. The red solid line is the summation of the best fits of signal and background. The contribution from isotope $\rm ^{68}Ga~(M)$ is too small to be seen in this plot.}
    \label{fig_inte_fit}
\end{figure}

\begin{figure}[!htbp]
\includegraphics[width=\linewidth]{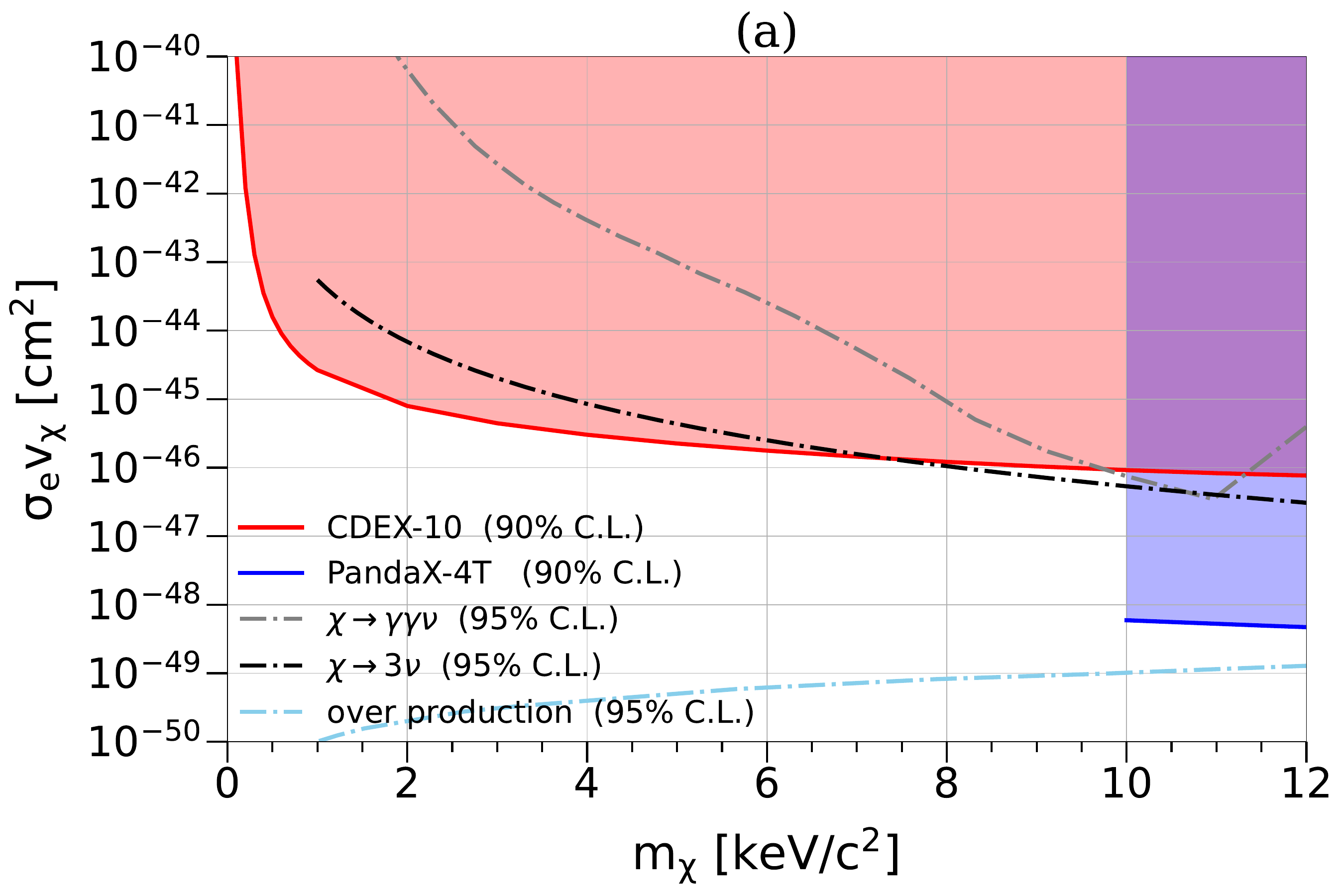}
\includegraphics[width=\linewidth]{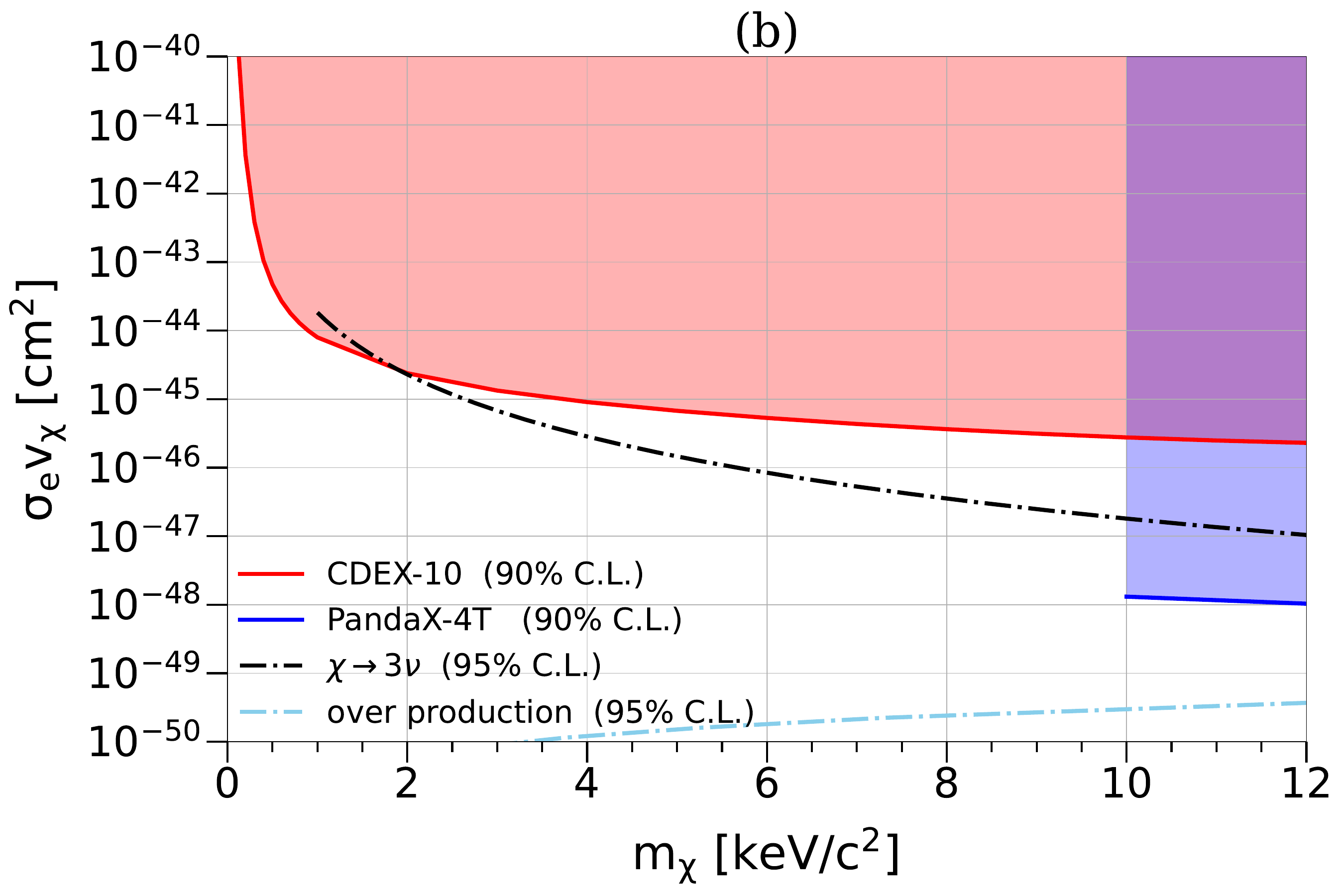}
\caption{The 90\% confidence level (CL) exclusion limits on $\sigma_e v_\chi$ of the fermionic DM absorbed by electron target for the (a) axial-vector operator and (b) vector operator from CDEX-10 is shown in red. Results from PandaX-4T~\cite{FDEA_3_PandaX} (blue) are superimposed. The exclusion regions of direct detection experiments are shown with shade. Results of DM invisible decay($\chi \rightarrow 3\nu$) and overproduction for both operators and DM visible decay($\chi\rightarrow\gamma\gamma\nu$) for the axial-vector operator are also shown (dashed dot line)~\cite{FDEA_1_GSF,CMB_Plank}.}
\label{fig_limits}
\end{figure}

The DM signal is fitted using the minimum-$\chi^2$ method~\cite{chisquare} for a given DM mass $m_\chi$ with the experiment data corresponding to the energy range from 0.16 keVee to 2.16 keVee. An example of best fit of $m_\chi$ = 12 keV/$c^2$ in the vector interaction case is illustrated in Fig.~\ref{fig_inte_fit}. Since no significant DM signals over the background are observed, an upper limit with 90\% confidence level (CL) is derived using the Feldman-Cousins method~\cite{FC_Method}. Figure~\ref{fig_limits} (a) and (b) shows the exclusion results of vector and axial-vector operators, respectively. We use natural units when reporting cross section results. The upper limit on the cross section is set to be $\rm 6.8\times10^{-46}~cm^2$ for vector interaction, and $\rm 2.3\times10^{-46}~cm^2$ for axial-vector interaction at DM mass of 5 keV/$c^2$. For the keV-scale electron-absorption processes studied here, the typical momentum transfer is $\mathcal{O}\rm(keV)$. As long as the mediator mass much larger than momentum transfer, the interaction is reliably described by the dimension-6 EFT operator, where, e.g., mediator mass is $\mathcal{O} \rm(MeV)$. This condition is easily satisfied in the dark-photon UV completion, and the EFT framework remains valid for our analysis.

Since the DM is unstable, it can decay into light SM particles. The dominant visible mode of the DM decay for axial-vector operator $\chi\rightarrow\gamma\gamma\nu$ would contribute to X(gamma)-ray observations around the Earth, which will be limited by X(gamma)-ray fluxes from astrophysics observations~\cite{FDEA_1_GSF}. For the vector operator, due to the gauge symmetry and quantum electrodynamics (QED) charge conjugation symmetry, $\chi\rightarrow\gamma\nu$ and $\chi\rightarrow\gamma\gamma\nu$ cannot arise. The dominant visible decay channel is $\chi\rightarrow\gamma\gamma\gamma\nu$, which is also limited by X(gamma)-ray fluxes from astrophysics observation. The constraints from $\chi \rightarrow 3\nu$ and overproduction~\cite{FDEA_1_GSF} for both operators and $\chi\rightarrow\gamma\gamma\nu$ for the axial-vector operator are depicted in Fig.~\ref{fig_limits}. The constraint from $\chi\rightarrow\gamma\gamma\gamma\nu$ for vector operator is weaker than the top edge of Fig.~\ref{fig_limits}, which is not shown.

Among the direct detection experiments, results from PandaX-4T using a liquid xenon time projection chamber~\cite{FDEA_3_PandaX} presents the most stringent constraints on the fermionic DM in the mass range of 35--55 (25--45) keV/$c^2$ for the vector (axial-vector) operator compared with constraints from x-ray satellites and large-scale observations. We place new constraints on the fermionic DM absorbed by electron targets for a lower DM mass below 10 keV/$c^2$ because of the low detector threshold of Ge semiconductor detectors used in CDEX experiment. 

\emph{Summary.}— 
In this paper, we report the constraints on $\sigma_e v_\chi$ for fermionic DM absorption by electron targets using data from the CDEX-10 experiment with an exposure of 205.4 kg$\cdot$day. Comparing with $\chi+A\rightarrow\nu+A$, which is only sensitive to DM in the MeV/$c^2$ mass scale, $\chi + e^- \rightarrow \nu + e^-$ can explore lighter DM for few keV/$c^2$. The expected DM signals of the DM mass range of 0.1--10 keV/$c^2$ are analyzed together with the experiment spectrum from 160 eVee to 2.16 keVee using a $\chi^2$ analysis method. With the analysis threshold of 160 eVee, we present new constraints on $\sigma_e v_\chi$ of the fermionic DM absorbed by electron targets in the DM range of 0.1--10 keV/$c^2$ for vector and axial-vector operators. Our limits reach the current lowest fermionic DM mass in direct detection experiments.

This work was supported by the National Key Research and Development Program of China (Grants No. 2023YFA1607101 and No. 2022YFA1605000) and the National Natural Science Foundation of China (Grants No. 12322511, No. 12175112, and No. 123B2087). We would like to thank CJPL and its staff for hosting and supporting the CDEX project. CJPL is jointly operated by Tsinghua University and Yalong River Hydropower Development Company.

\bibliography{DMAbsorption}

 \newcommand{\noop}[1]{}
\begin{thebibliography}{67}%
\makeatletter
\providecommand \@ifxundefined [1]{%
 \@ifx{#1\undefined}
}%
\providecommand \@ifnum [1]{%
 \ifnum #1\expandafter \@firstoftwo
 \else \expandafter \@secondoftwo
 \fi
}%
\providecommand \@ifx [1]{%
 \ifx #1\expandafter \@firstoftwo
 \else \expandafter \@secondoftwo
 \fi
}%
\providecommand \natexlab [1]{#1}%
\providecommand \enquote  [1]{``#1''}%
\providecommand \bibnamefont  [1]{#1}%
\providecommand \bibfnamefont [1]{#1}%
\providecommand \citenamefont [1]{#1}%
\providecommand \href@noop [0]{\@secondoftwo}%
\providecommand \href [0]{\begingroup \@sanitize@url \@href}%
\providecommand \@href[1]{\@@startlink{#1}\@@href}%
\providecommand \@@href[1]{\endgroup#1\@@endlink}%
\providecommand \@sanitize@url [0]{\catcode `\\12\catcode `\$12\catcode
  `\&12\catcode `\#12\catcode `\^12\catcode `\_12\catcode `\%12\relax}%
\providecommand \@@startlink[1]{}%
\providecommand \@@endlink[0]{}%
\providecommand \url  [0]{\begingroup\@sanitize@url \@url }%
\providecommand \@url [1]{\endgroup\@href {#1}{\urlprefix }}%
\providecommand \urlprefix  [0]{URL }%
\providecommand \Eprint [0]{\href }%
\providecommand \doibase [0]{http://dx.doi.org/}%
\providecommand \selectlanguage [0]{\@gobble}%
\providecommand \bibinfo  [0]{\@secondoftwo}%
\providecommand \bibfield  [0]{\@secondoftwo}%
\providecommand \translation [1]{[#1]}%
\providecommand \BibitemOpen [0]{}%
\providecommand \bibitemStop [0]{}%
\providecommand \bibitemNoStop [0]{.\EOS\space}%
\providecommand \EOS [0]{\spacefactor3000\relax}%
\providecommand \BibitemShut  [1]{\csname bibitem#1\endcsname}%
\let\auto@bib@innerbib\@empty
\bibitem [{\citenamefont {Cho}(2013)}]{DM_Universe}%
  \BibitemOpen
  \bibfield  {author} {\bibinfo {author} {\bibfnamefont {A.}~\bibnamefont
  {Cho}},\ }\href {\doibase 10.1126/science.339.6127.1513} {\bibfield
  {journal} {\bibinfo  {journal} {Science}\ }\textbf {\bibinfo {volume}
  {339}},\ \bibinfo {pages} {1513} (\bibinfo {year} {2013})}\BibitemShut
  {NoStop}%
\bibitem [{\citenamefont {Aprile}\ \emph {et~al.}(2023)\citenamefont {Aprile}
  \emph {et~al.}}]{XENONnT_2023}%
  \BibitemOpen
  \bibfield  {author} {\bibinfo {author} {\bibfnamefont {E.}~\bibnamefont
  {Aprile}} \emph {et~al.} (\bibinfo {collaboration} {XENON Collaboration}),\
  }\href {\doibase 10.1103/PhysRevLett.131.041003} {\bibfield  {journal}
  {\bibinfo  {journal} {Phys. Rev. Lett.}\ }\textbf {\bibinfo {volume} {131}},\
  \bibinfo {pages} {041003} (\bibinfo {year} {2023})}\BibitemShut {NoStop}%
\bibitem [{\citenamefont {Aalbers}\ \emph {et~al.}(2023)\citenamefont {Aalbers}
  \emph {et~al.}}]{LZ_2023}%
  \BibitemOpen
  \bibfield  {author} {\bibinfo {author} {\bibfnamefont {J.}~\bibnamefont
  {Aalbers}} \emph {et~al.} (\bibinfo {collaboration} {LUX-ZEPLIN
  Collaboration}),\ }\href {\doibase 10.1103/PhysRevLett.131.041002} {\bibfield
   {journal} {\bibinfo  {journal} {Phys. Rev. Lett.}\ }\textbf {\bibinfo
  {volume} {131}},\ \bibinfo {pages} {041002} (\bibinfo {year}
  {2023})}\BibitemShut {NoStop}%
\bibitem [{\citenamefont {Li}\ \emph {et~al.}(2023)\citenamefont {Li} \emph
  {et~al.}}]{PandaX-4T_2023}%
  \BibitemOpen
  \bibfield  {author} {\bibinfo {author} {\bibfnamefont {S.}~\bibnamefont {Li}}
  \emph {et~al.} (\bibinfo {collaboration} {PandaX Collaboration}),\ }\href
  {\doibase 10.1103/PhysRevLett.130.261001} {\bibfield  {journal} {\bibinfo
  {journal} {Phys. Rev. Lett.}\ }\textbf {\bibinfo {volume} {130}},\ \bibinfo
  {pages} {261001} (\bibinfo {year} {2023})}\BibitemShut {NoStop}%
\bibitem [{\citenamefont {Agnes}\ \emph
  {et~al.}(2018{\natexlab{a}})\citenamefont {Agnes} \emph {et~al.}}]{DarkSide}%
  \BibitemOpen
  \bibfield  {author} {\bibinfo {author} {\bibfnamefont {P.}~\bibnamefont
  {Agnes}} \emph {et~al.} (\bibinfo {collaboration} {DarkSide Collaboration}),\
  }\href {\doibase 10.1103/PhysRevLett.121.081307} {\bibfield  {journal}
  {\bibinfo  {journal} {Phys. Rev. Lett.}\ }\textbf {\bibinfo {volume} {121}},\
  \bibinfo {pages} {081307} (\bibinfo {year} {2018}{\natexlab{a}})}\BibitemShut
  {NoStop}%
\bibitem [{\citenamefont {Agnese}\ \emph {et~al.}(2018)\citenamefont {Agnese}
  \emph {et~al.}}]{SuperCDMS}%
  \BibitemOpen
  \bibfield  {author} {\bibinfo {author} {\bibfnamefont {R.}~\bibnamefont
  {Agnese}} \emph {et~al.} (\bibinfo {collaboration} {SuperCDMS
  Collaboration}),\ }\href {\doibase 10.1103/PhysRevD.97.022002} {\bibfield
  {journal} {\bibinfo  {journal} {Phys. Rev. D}\ }\textbf {\bibinfo {volume}
  {97}},\ \bibinfo {pages} {022002} (\bibinfo {year} {2018})}\BibitemShut
  {NoStop}%
\bibitem [{\citenamefont {Liu}\ \emph {et~al.}(2014)\citenamefont {Liu} \emph
  {et~al.}}]{CDEX_1_LSK}%
  \BibitemOpen
  \bibfield  {author} {\bibinfo {author} {\bibfnamefont {S.~K.}\ \bibnamefont
  {Liu}} \emph {et~al.} (\bibinfo {collaboration} {CDEX Collaboration}),\
  }\href {\doibase 10.1103/PhysRevD.90.032003} {\bibfield  {journal} {\bibinfo
  {journal} {Phys. Rev. D}\ }\textbf {\bibinfo {volume} {90}},\ \bibinfo
  {pages} {032003} (\bibinfo {year} {2014})}\BibitemShut {NoStop}%
\bibitem [{\citenamefont {Zhao}\ \emph {et~al.}(2013)\citenamefont {Zhao} \emph
  {et~al.}}]{CDEX_2_ZW}%
  \BibitemOpen
  \bibfield  {author} {\bibinfo {author} {\bibfnamefont {W.}~\bibnamefont
  {Zhao}} \emph {et~al.} (\bibinfo {collaboration} {CDEX Collaboration}),\
  }\href {\doibase 10.1103/PhysRevD.88.052004} {\bibfield  {journal} {\bibinfo
  {journal} {Phys. Rev. D}\ }\textbf {\bibinfo {volume} {88}},\ \bibinfo
  {pages} {052004} (\bibinfo {year} {2013})}\BibitemShut {NoStop}%
\bibitem [{\citenamefont {Yue}\ \emph {et~al.}(2014)\citenamefont {Yue} \emph
  {et~al.}}]{CDEX_3_YQ}%
  \BibitemOpen
  \bibfield  {author} {\bibinfo {author} {\bibfnamefont {Q.}~\bibnamefont
  {Yue}} \emph {et~al.} (\bibinfo {collaboration} {CDEX Collaboration}),\
  }\href {\doibase 10.1103/PhysRevD.90.091701} {\bibfield  {journal} {\bibinfo
  {journal} {Phys. Rev. D}\ }\textbf {\bibinfo {volume} {90}},\ \bibinfo
  {pages} {091701} (\bibinfo {year} {2014})}\BibitemShut {NoStop}%
\bibitem [{\citenamefont {Zhao}\ \emph {et~al.}(2016)\citenamefont {Zhao} \emph
  {et~al.}}]{CDEX_4_ZW}%
  \BibitemOpen
  \bibfield  {author} {\bibinfo {author} {\bibfnamefont {W.}~\bibnamefont
  {Zhao}} \emph {et~al.} (\bibinfo {collaboration} {CDEX Collaboration}),\
  }\href {\doibase 10.1103/PhysRevD.93.092003} {\bibfield  {journal} {\bibinfo
  {journal} {Phys. Rev. D}\ }\textbf {\bibinfo {volume} {93}},\ \bibinfo
  {pages} {092003} (\bibinfo {year} {2016})}\BibitemShut {NoStop}%
\bibitem [{\citenamefont {Yang}\ \emph
  {et~al.}(2018{\natexlab{a}})\citenamefont {Yang} \emph
  {et~al.}}]{CDEX_5_YLT}%
  \BibitemOpen
  \bibfield  {author} {\bibinfo {author} {\bibfnamefont {L.~T.}\ \bibnamefont
  {Yang}} \emph {et~al.} (\bibinfo {collaboration} {CDEX Collaboration}),\
  }\href {\doibase 10.1088/1674-1137/42/2/023002} {\bibfield  {journal}
  {\bibinfo  {journal} {Chin. Phys. C}\ }\textbf {\bibinfo {volume} {42}},\
  \bibinfo {eid} {023002} (\bibinfo {year} {2018}{\natexlab{a}})}\BibitemShut
  {NoStop}%
\bibitem [{\citenamefont {Jiang}\ \emph {et~al.}(2018)\citenamefont {Jiang}
  \emph {et~al.}}]{CDEX_6_JH}%
  \BibitemOpen
  \bibfield  {author} {\bibinfo {author} {\bibfnamefont {H.}~\bibnamefont
  {Jiang}} \emph {et~al.} (\bibinfo {collaboration} {CDEX Collaboration}),\
  }\href {\doibase 10.1103/PhysRevLett.120.241301} {\bibfield  {journal}
  {\bibinfo  {journal} {Phys. Rev. Lett.}\ }\textbf {\bibinfo {volume} {120}},\
  \bibinfo {pages} {241301} (\bibinfo {year} {2018})}\BibitemShut {NoStop}%
\bibitem [{\citenamefont {Jiang}\ \emph {et~al.}(2019)\citenamefont {Jiang}
  \emph {et~al.}}]{CDEX_7_JH}%
  \BibitemOpen
  \bibfield  {author} {\bibinfo {author} {\bibfnamefont {H.}~\bibnamefont
  {Jiang}} \emph {et~al.} (\bibinfo {collaboration} {CDEX Collaboration}),\
  }\href {\doibase 10.1103/PhysRevLett.120.241301} {\bibfield  {journal}
  {\bibinfo  {journal} {Sci. China Phys. Mech. Astron.}\ }\textbf {\bibinfo
  {volume} {62}},\ \bibinfo {pages} {031012} (\bibinfo {year}
  {2019})}\BibitemShut {NoStop}%
\bibitem [{\citenamefont {Yang}\ \emph {et~al.}(2019)\citenamefont {Yang} \emph
  {et~al.}}]{CDEX_8_YLT}%
  \BibitemOpen
  \bibfield  {author} {\bibinfo {author} {\bibfnamefont {L.~T.}\ \bibnamefont
  {Yang}} \emph {et~al.} (\bibinfo {collaboration} {CDEX Collaboration}),\
  }\href {\doibase 10.1103/PhysRevLett.123.221301} {\bibfield  {journal}
  {\bibinfo  {journal} {Phys. Rev. Lett.}\ }\textbf {\bibinfo {volume} {123}},\
  \bibinfo {pages} {221301} (\bibinfo {year} {2019})}\BibitemShut {NoStop}%
\bibitem [{\citenamefont {Liu}\ \emph {et~al.}(2019)\citenamefont {Liu} \emph
  {et~al.}}]{CDEX_9_LZZ}%
  \BibitemOpen
  \bibfield  {author} {\bibinfo {author} {\bibfnamefont {Z.~Z.}\ \bibnamefont
  {Liu}} \emph {et~al.} (\bibinfo {collaboration} {CDEX Collaboration}),\
  }\href {\doibase 10.1103/PhysRevLett.123.161301} {\bibfield  {journal}
  {\bibinfo  {journal} {Phys. Rev. Lett.}\ }\textbf {\bibinfo {volume} {123}},\
  \bibinfo {pages} {161301} (\bibinfo {year} {2019})}\BibitemShut {NoStop}%
\bibitem [{\citenamefont {Akerib}\ \emph {et~al.}(2019)\citenamefont {Akerib}
  \emph {et~al.}}]{Migdal_2_LUX}%
  \BibitemOpen
  \bibfield  {author} {\bibinfo {author} {\bibfnamefont {D.~S.}\ \bibnamefont
  {Akerib}} \emph {et~al.} (\bibinfo {collaboration} {LUX Collaboration}),\
  }\href {\doibase 10.1103/PhysRevLett.122.131301} {\bibfield  {journal}
  {\bibinfo  {journal} {Phys. Rev. Lett.}\ }\textbf {\bibinfo {volume} {122}},\
  \bibinfo {pages} {131301} (\bibinfo {year} {2019})}\BibitemShut {NoStop}%
\bibitem [{\citenamefont {Liu}\ \emph {et~al.}(2022)\citenamefont {Liu} \emph
  {et~al.}}]{Migdal_4_LZZ}%
  \BibitemOpen
  \bibfield  {author} {\bibinfo {author} {\bibfnamefont {Z.~Z.}\ \bibnamefont
  {Liu}} \emph {et~al.} (\bibinfo {collaboration} {CDEX Collaboration}),\
  }\href {\doibase 10.1103/PhysRevD.105.052005} {\bibfield  {journal} {\bibinfo
   {journal} {Phys. Rev. D}\ }\textbf {\bibinfo {volume} {105}},\ \bibinfo
  {pages} {052005} (\bibinfo {year} {2022})}\BibitemShut {NoStop}%
\bibitem [{\citenamefont {Angloher}\ \emph {et~al.}(2017)\citenamefont
  {Angloher} \emph {et~al.}}]{Migdal_3_CRESST}%
  \BibitemOpen
  \bibfield  {author} {\bibinfo {author} {\bibfnamefont {G.}~\bibnamefont
  {Angloher}} \emph {et~al.} (\bibinfo {collaboration} {CRESST
  Collaboration}),\ }\href {\doibase 10.1140/epjc/s10052-017-5223-9} {\bibfield
   {journal} {\bibinfo  {journal} {Eur. Phys. J. C}\ }\textbf {\bibinfo
  {volume} {77}},\ \bibinfo {pages} {637} (\bibinfo {year} {2017})}\BibitemShut
  {NoStop}%
\bibitem [{\citenamefont {Dror}\ \emph
  {et~al.}(2020{\natexlab{a}})\citenamefont {Dror}, \citenamefont {Elor},\ and\
  \citenamefont {McGehee}}]{FDNA_1_Dror}%
  \BibitemOpen
  \bibfield  {author} {\bibinfo {author} {\bibfnamefont {J.~A.}\ \bibnamefont
  {Dror}}, \bibinfo {author} {\bibfnamefont {G.}~\bibnamefont {Elor}}, \ and\
  \bibinfo {author} {\bibfnamefont {R.}~\bibnamefont {McGehee}},\ }\href
  {\doibase 10.1103/PhysRevLett.124.181301} {\bibfield  {journal} {\bibinfo
  {journal} {Phys. Rev. Lett.}\ }\textbf {\bibinfo {volume} {124}},\ \bibinfo
  {pages} {181301} (\bibinfo {year} {2020}{\natexlab{a}})}\BibitemShut
  {NoStop}%
\bibitem [{\citenamefont {Dror}\ \emph
  {et~al.}(2020{\natexlab{b}})\citenamefont {Dror}, \citenamefont {Elor},\ and\
  \citenamefont {McGehee}}]{FDNA_2_Dror}%
  \BibitemOpen
  \bibfield  {author} {\bibinfo {author} {\bibfnamefont {J.~A.}\ \bibnamefont
  {Dror}}, \bibinfo {author} {\bibfnamefont {G.}~\bibnamefont {Elor}}, \ and\
  \bibinfo {author} {\bibfnamefont {R.}~\bibnamefont {McGehee}},\ }\href
  {\doibase 10.1007/JHEP02(2020)134} {\bibfield  {journal} {\bibinfo  {journal}
  {J. High Energy Phys.}\ }\textbf {\bibinfo {volume} {02}},\ \bibinfo {pages}
  {134} (\bibinfo {year} {2020}{\natexlab{b}})}\BibitemShut {NoStop}%
\bibitem [{\citenamefont {Gu}\ \emph {et~al.}(2022)\citenamefont {Gu} \emph
  {et~al.}}]{FDNA_3_PandaX}%
  \BibitemOpen
  \bibfield  {author} {\bibinfo {author} {\bibfnamefont {L.}~\bibnamefont {Gu}}
  \emph {et~al.} (\bibinfo {collaboration} {PandaX Collaboration}),\ }\href
  {\doibase 10.1103/PhysRevLett.129.161803} {\bibfield  {journal} {\bibinfo
  {journal} {Phys. Rev. Lett.}\ }\textbf {\bibinfo {volume} {129}},\ \bibinfo
  {pages} {161803} (\bibinfo {year} {2022})}\BibitemShut {NoStop}%
\bibitem [{\citenamefont {Arnquist}\ \emph {et~al.}(2024)\citenamefont
  {Arnquist} \emph {et~al.}}]{FDNA_4_MJD}%
  \BibitemOpen
  \bibfield  {author} {\bibinfo {author} {\bibfnamefont {I.~J.}\ \bibnamefont
  {Arnquist}} \emph {et~al.} (\bibinfo {collaboration} {Majorana
  Collaboration}),\ }\href {\doibase 10.1103/PhysRevLett.132.041001} {\bibfield
   {journal} {\bibinfo  {journal} {Phys. Rev. Lett.}\ }\textbf {\bibinfo
  {volume} {132}},\ \bibinfo {pages} {041001} (\bibinfo {year}
  {2024})}\BibitemShut {NoStop}%
\bibitem [{\citenamefont {Dai}\ \emph {et~al.}(2022)\citenamefont {Dai} \emph
  {et~al.}}]{FDNA_5_DWH}%
  \BibitemOpen
  \bibfield  {author} {\bibinfo {author} {\bibfnamefont {W.~H.}\ \bibnamefont
  {Dai}} \emph {et~al.} (\bibinfo {collaboration} {CDEX Collaboration}),\
  }\href {\doibase 10.1103/PhysRevLett.129.221802} {\bibfield  {journal}
  {\bibinfo  {journal} {Phys. Rev. Lett.}\ }\textbf {\bibinfo {volume} {129}},\
  \bibinfo {pages} {221802} (\bibinfo {year} {2022})}\BibitemShut {NoStop}%
\bibitem [{\citenamefont {Chao}\ \emph {et~al.}(2023)\citenamefont {Chao},
  \citenamefont {Jin},\ and\ \citenamefont {Peng}}]{3to2_CW}%
  \BibitemOpen
  \bibfield  {author} {\bibinfo {author} {\bibfnamefont {W.}~\bibnamefont
  {Chao}}, \bibinfo {author} {\bibfnamefont {M.}~\bibnamefont {Jin}}, \ and\
  \bibinfo {author} {\bibfnamefont {Y.~Q.}\ \bibnamefont {Peng}},\ }\href
  {\doibase 10.1103/PhysRevD.107.093009} {\bibfield  {journal} {\bibinfo
  {journal} {Phys. Rev. D}\ }\textbf {\bibinfo {volume} {107}},\ \bibinfo
  {pages} {093009} (\bibinfo {year} {2023})}\BibitemShut {NoStop}%
\bibitem [{\citenamefont {Li}\ \emph {et~al.}(2022)\citenamefont {Li},
  \citenamefont {Liao},\ and\ \citenamefont {Zhang}}]{FDMD_Li}%
  \BibitemOpen
  \bibfield  {author} {\bibinfo {author} {\bibfnamefont {T.}~\bibnamefont
  {Li}}, \bibinfo {author} {\bibfnamefont {J.}~\bibnamefont {Liao}}, \ and\
  \bibinfo {author} {\bibfnamefont {R.~J.}\ \bibnamefont {Zhang}},\ }\href
  {\doibase 10.1007/JHEP05(2022)071} {\bibfield  {journal} {\bibinfo  {journal}
  {J. High Energy Phys.}\ }\textbf {\bibinfo {volume} {05}},\ \bibinfo {pages}
  {071} (\bibinfo {year} {2022})}\BibitemShut {NoStop}%
\bibitem [{\citenamefont {Aguilar-Arevalo}\ \emph {et~al.}(2019)\citenamefont
  {Aguilar-Arevalo} \emph {et~al.}}]{chi-e_DAMIC}%
  \BibitemOpen
  \bibfield  {author} {\bibinfo {author} {\bibfnamefont {A.}~\bibnamefont
  {Aguilar-Arevalo}} \emph {et~al.} (\bibinfo {collaboration} {DAMIC
  Collaboration}),\ }\href {\doibase 10.1103/PhysRevLett.123.181802} {\bibfield
   {journal} {\bibinfo  {journal} {Phys. Rev. Lett.}\ }\textbf {\bibinfo
  {volume} {123}},\ \bibinfo {pages} {181802} (\bibinfo {year}
  {2019})}\BibitemShut {NoStop}%
\bibitem [{\citenamefont {Agnes}\ \emph
  {et~al.}(2018{\natexlab{b}})\citenamefont {Agnes} \emph
  {et~al.}}]{chi-e_DarkSide}%
  \BibitemOpen
  \bibfield  {author} {\bibinfo {author} {\bibfnamefont {P.}~\bibnamefont
  {Agnes}} \emph {et~al.} (\bibinfo {collaboration} {DarkSide Collaboration}),\
  }\href {\doibase 10.1103/PhysRevLett.121.111303} {\bibfield  {journal}
  {\bibinfo  {journal} {Phys. Rev. Lett.}\ }\textbf {\bibinfo {volume} {121}},\
  \bibinfo {pages} {111303} (\bibinfo {year} {2018}{\natexlab{b}})}\BibitemShut
  {NoStop}%
\bibitem [{\citenamefont {Arnaud}\ \emph {et~al.}(2020)\citenamefont {Arnaud}
  \emph {et~al.}}]{chi-e_EDELWEISS}%
  \BibitemOpen
  \bibfield  {author} {\bibinfo {author} {\bibfnamefont {Q.}~\bibnamefont
  {Arnaud}} \emph {et~al.} (\bibinfo {collaboration} {EDELWEISS
  Collaboration}),\ }\href {\doibase 10.1103/PhysRevLett.125.141301} {\bibfield
   {journal} {\bibinfo  {journal} {Phys. Rev. Lett.}\ }\textbf {\bibinfo
  {volume} {125}},\ \bibinfo {pages} {141301} (\bibinfo {year}
  {2020})}\BibitemShut {NoStop}%
\bibitem [{\citenamefont {Cheng}\ \emph {et~al.}(2021)\citenamefont {Cheng}
  \emph {et~al.}}]{chi-e_PandaX}%
  \BibitemOpen
  \bibfield  {author} {\bibinfo {author} {\bibfnamefont {C.}~\bibnamefont
  {Cheng}} \emph {et~al.} (\bibinfo {collaboration} {PandaX-II
  Collaboration}),\ }\href {\doibase 10.1103/PhysRevLett.126.211803} {\bibfield
   {journal} {\bibinfo  {journal} {Phys. Rev. Lett.}\ }\textbf {\bibinfo
  {volume} {126}},\ \bibinfo {pages} {211803} (\bibinfo {year}
  {2021})}\BibitemShut {NoStop}%
\bibitem [{\citenamefont {Barak}\ \emph {et~al.}(2020)\citenamefont {Barak}
  \emph {et~al.}}]{chi-e_SENSEI}%
  \BibitemOpen
  \bibfield  {author} {\bibinfo {author} {\bibfnamefont {L.}~\bibnamefont
  {Barak}} \emph {et~al.} (\bibinfo {collaboration} {SENSEI Collaboration}),\
  }\href {\doibase 10.1103/PhysRevLett.125.171802} {\bibfield  {journal}
  {\bibinfo  {journal} {Phys. Rev. Lett.}\ }\textbf {\bibinfo {volume} {125}},\
  \bibinfo {pages} {171802} (\bibinfo {year} {2020})}\BibitemShut {NoStop}%
\bibitem [{\citenamefont {Amaral}\ \emph {et~al.}(2020)\citenamefont {Amaral}
  \emph {et~al.}}]{chi-e_SuperCDMS}%
  \BibitemOpen
  \bibfield  {author} {\bibinfo {author} {\bibfnamefont {D.~W.}\ \bibnamefont
  {Amaral}} \emph {et~al.},\ }\href {\doibase 10.1103/PhysRevD.102.091101}
  {\bibfield  {journal} {\bibinfo  {journal} {Phys. Rev. D}\ }\textbf {\bibinfo
  {volume} {102}},\ \bibinfo {pages} {091101} (\bibinfo {year}
  {2020})}\BibitemShut {NoStop}%
\bibitem [{\citenamefont {Essig}\ \emph {et~al.}(2017)\citenamefont {Essig},
  \citenamefont {Volansky},\ and\ \citenamefont {Yu}}]{chi-e_XENON}%
  \BibitemOpen
  \bibfield  {author} {\bibinfo {author} {\bibfnamefont {R.}~\bibnamefont
  {Essig}}, \bibinfo {author} {\bibfnamefont {T.}~\bibnamefont {Volansky}}, \
  and\ \bibinfo {author} {\bibfnamefont {T.-T.}\ \bibnamefont {Yu}},\ }\href
  {\doibase 10.1103/PhysRevD.96.043017} {\bibfield  {journal} {\bibinfo
  {journal} {Phys. Rev. D}\ }\textbf {\bibinfo {volume} {96}},\ \bibinfo
  {pages} {043017} (\bibinfo {year} {2017})}\BibitemShut {NoStop}%
\bibitem [{\citenamefont {Aprile}\ \emph {et~al.}(2019)\citenamefont {Aprile}
  \emph {et~al.}}]{chi-e_XENON_2}%
  \BibitemOpen
  \bibfield  {author} {\bibinfo {author} {\bibfnamefont {E.}~\bibnamefont
  {Aprile}} \emph {et~al.} (\bibinfo {collaboration} {XENON Collaboration}),\
  }\href {\doibase 10.1103/PhysRevLett.123.251801} {\bibfield  {journal}
  {\bibinfo  {journal} {Phys. Rev. Lett.}\ }\textbf {\bibinfo {volume} {123}},\
  \bibinfo {pages} {251801} (\bibinfo {year} {2019})}\BibitemShut {NoStop}%
\bibitem [{\citenamefont {Zhang}\ \emph
  {et~al.}(2022{\natexlab{a}})\citenamefont {Zhang} \emph {et~al.}}]{CDEX_ZZY}%
  \BibitemOpen
  \bibfield  {author} {\bibinfo {author} {\bibfnamefont {Z.~Y.}\ \bibnamefont
  {Zhang}} \emph {et~al.} (\bibinfo {collaboration} {CDEX Collaboration}),\
  }\href {\doibase 10.1103/PhysRevLett.129.221301} {\bibfield  {journal}
  {\bibinfo  {journal} {Phys. Rev. Lett.}\ }\textbf {\bibinfo {volume} {129}},\
  \bibinfo {pages} {221301} (\bibinfo {year} {2022}{\natexlab{a}})}\BibitemShut
  {NoStop}%
\bibitem [{\citenamefont {Bringmann}\ and\ \citenamefont
  {Pospelov}(2019)}]{CRDM_25}%
  \BibitemOpen
  \bibfield  {author} {\bibinfo {author} {\bibfnamefont {T.}~\bibnamefont
  {Bringmann}}\ and\ \bibinfo {author} {\bibfnamefont {M.}~\bibnamefont
  {Pospelov}},\ }\href {\doibase 10.1103/PhysRevLett.122.171801} {\bibfield
  {journal} {\bibinfo  {journal} {Phys. Rev. Lett.}\ }\textbf {\bibinfo
  {volume} {122}},\ \bibinfo {pages} {171801} (\bibinfo {year}
  {2019})}\BibitemShut {NoStop}%
\bibitem [{\citenamefont {Lei}\ \emph {et~al.}(2022)\citenamefont {Lei},
  \citenamefont {Tang},\ and\ \citenamefont {Zhang}}]{CRDM_26}%
  \BibitemOpen
  \bibfield  {author} {\bibinfo {author} {\bibfnamefont {Z.~H.}\ \bibnamefont
  {Lei}}, \bibinfo {author} {\bibfnamefont {J.}~\bibnamefont {Tang}}, \ and\
  \bibinfo {author} {\bibfnamefont {B.~L.}\ \bibnamefont {Zhang}},\ }\href
  {\doibase 10.1088/1674-1137/ac68da} {\bibfield  {journal} {\bibinfo
  {journal} {Chin. Phys. C}\ }\textbf {\bibinfo {volume} {46}},\ \bibinfo
  {pages} {085103} (\bibinfo {year} {2022})}\BibitemShut {NoStop}%
\bibitem [{\citenamefont {Xia}\ \emph {et~al.}(2021)\citenamefont {Xia},
  \citenamefont {Xu},\ and\ \citenamefont {Zhou}}]{CRDM_27}%
  \BibitemOpen
  \bibfield  {author} {\bibinfo {author} {\bibfnamefont {C.}~\bibnamefont
  {Xia}}, \bibinfo {author} {\bibfnamefont {Y.~H.}\ \bibnamefont {Xu}}, \ and\
  \bibinfo {author} {\bibfnamefont {Y.~F.}\ \bibnamefont {Zhou}},\ }\href
  {\doibase 10.1016/j.nuclphysb.2021.115470} {\bibfield  {journal} {\bibinfo
  {journal} {Nucl. Phys. B}\ }\textbf {\bibinfo {volume} {969}},\ \bibinfo
  {pages} {115470} (\bibinfo {year} {2021})}\BibitemShut {NoStop}%
\bibitem [{\citenamefont {Cappiello}\ and\ \citenamefont
  {Beacom}(2019)}]{CRDM_28}%
  \BibitemOpen
  \bibfield  {author} {\bibinfo {author} {\bibfnamefont {C.~V.}\ \bibnamefont
  {Cappiello}}\ and\ \bibinfo {author} {\bibfnamefont {J.~F.}\ \bibnamefont
  {Beacom}},\ }\href {\doibase 10.1103/PhysRevD.100.103011} {\bibfield
  {journal} {\bibinfo  {journal} {Phys. Rev. D}\ }\textbf {\bibinfo {volume}
  {100}},\ \bibinfo {pages} {103011} (\bibinfo {year} {2019})}\BibitemShut
  {NoStop}%
\bibitem [{\citenamefont {Ema}\ \emph {et~al.}(2019)\citenamefont {Ema},
  \citenamefont {Sala},\ and\ \citenamefont {Sato}}]{CRDM_29}%
  \BibitemOpen
  \bibfield  {author} {\bibinfo {author} {\bibfnamefont {Y.}~\bibnamefont
  {Ema}}, \bibinfo {author} {\bibfnamefont {F.}~\bibnamefont {Sala}}, \ and\
  \bibinfo {author} {\bibfnamefont {R.}~\bibnamefont {Sato}},\ }\href {\doibase
  10.1103/PhysRevLett.122.181802} {\bibfield  {journal} {\bibinfo  {journal}
  {Phys. Rev. Lett.}\ }\textbf {\bibinfo {volume} {122}},\ \bibinfo {pages}
  {181802} (\bibinfo {year} {2019})}\BibitemShut {NoStop}%
\bibitem [{\citenamefont {Dent}\ \emph {et~al.}(2020)\citenamefont {Dent},
  \citenamefont {Dutta}, \citenamefont {Newstead},\ and\ \citenamefont
  {Shoemaker}}]{CRDM_30}%
  \BibitemOpen
  \bibfield  {author} {\bibinfo {author} {\bibfnamefont {J.~B.}\ \bibnamefont
  {Dent}}, \bibinfo {author} {\bibfnamefont {B.}~\bibnamefont {Dutta}},
  \bibinfo {author} {\bibfnamefont {J.~L.}\ \bibnamefont {Newstead}}, \ and\
  \bibinfo {author} {\bibfnamefont {I.~M.}\ \bibnamefont {Shoemaker}},\ }\href
  {\doibase 10.1103/PhysRevD.101.116007} {\bibfield  {journal} {\bibinfo
  {journal} {Phys. Rev. D}\ }\textbf {\bibinfo {volume} {101}},\ \bibinfo
  {pages} {116007} (\bibinfo {year} {2020})}\BibitemShut {NoStop}%
\bibitem [{\citenamefont {Cui}\ \emph {et~al.}(2022)\citenamefont {Cui} \emph
  {et~al.}}]{CRDM_31}%
  \BibitemOpen
  \bibfield  {author} {\bibinfo {author} {\bibfnamefont {X.}~\bibnamefont
  {Cui}} \emph {et~al.} (\bibinfo {collaboration} {PandaX-II Collaboration}),\
  }\href {\doibase 10.1103/PhysRevLett.128.171801} {\bibfield  {journal}
  {\bibinfo  {journal} {Phys. Rev. Lett.}\ }\textbf {\bibinfo {volume} {128}},\
  \bibinfo {pages} {171801} (\bibinfo {year} {2022})}\BibitemShut {NoStop}%
\bibitem [{\citenamefont {Xu}\ \emph {et~al.}(2022)\citenamefont {Xu} \emph
  {et~al.}}]{CDEX_XR}%
  \BibitemOpen
  \bibfield  {author} {\bibinfo {author} {\bibfnamefont {R.}~\bibnamefont {Xu}}
  \emph {et~al.} (\bibinfo {collaboration} {CDEX Collaboration}),\ }\href
  {\doibase 10.1103/PhysRevD.106.052008} {\bibfield  {journal} {\bibinfo
  {journal} {Phys. Rev. D}\ }\textbf {\bibinfo {volume} {106}},\ \bibinfo
  {pages} {052008} (\bibinfo {year} {2022})}\BibitemShut {NoStop}%
\bibitem [{\citenamefont {Pospelov}\ \emph {et~al.}(2008)\citenamefont
  {Pospelov}, \citenamefont {Ritz},\ and\ \citenamefont
  {Voloshin}}]{Bosonic_1}%
  \BibitemOpen
  \bibfield  {author} {\bibinfo {author} {\bibfnamefont {M.}~\bibnamefont
  {Pospelov}}, \bibinfo {author} {\bibfnamefont {A.}~\bibnamefont {Ritz}}, \
  and\ \bibinfo {author} {\bibfnamefont {M.}~\bibnamefont {Voloshin}},\ }\href
  {\doibase 10.1103/PhysRevD.78.115012} {\bibfield  {journal} {\bibinfo
  {journal} {Phys. Rev. D}\ }\textbf {\bibinfo {volume} {78}},\ \bibinfo
  {pages} {115012} (\bibinfo {year} {2008})}\BibitemShut {NoStop}%
\bibitem [{\citenamefont {An}\ \emph {et~al.}(2015)\citenamefont {An},
  \citenamefont {Pospelov}, \citenamefont {Pradler},\ and\ \citenamefont
  {Ritz}}]{Bosonic_2}%
  \BibitemOpen
  \bibfield  {author} {\bibinfo {author} {\bibfnamefont {H.}~\bibnamefont
  {An}}, \bibinfo {author} {\bibfnamefont {M.}~\bibnamefont {Pospelov}},
  \bibinfo {author} {\bibfnamefont {J.}~\bibnamefont {Pradler}}, \ and\
  \bibinfo {author} {\bibfnamefont {A.}~\bibnamefont {Ritz}},\ }\href {\doibase
  10.1016/j.physletb.2015.06.018} {\bibfield  {journal} {\bibinfo  {journal}
  {Phys. Lett. B}\ }\textbf {\bibinfo {volume} {747}},\ \bibinfo {pages} {331}
  (\bibinfo {year} {2015})}\BibitemShut {NoStop}%
\bibitem [{\citenamefont {Hochberg}\ \emph {et~al.}(2016)\citenamefont
  {Hochberg}, \citenamefont {Lin},\ and\ \citenamefont {Zurek}}]{Bosonic_3}%
  \BibitemOpen
  \bibfield  {author} {\bibinfo {author} {\bibfnamefont {Y.}~\bibnamefont
  {Hochberg}}, \bibinfo {author} {\bibfnamefont {T.}~\bibnamefont {Lin}}, \
  and\ \bibinfo {author} {\bibfnamefont {K.~M.}\ \bibnamefont {Zurek}},\ }\href
  {\doibase 10.1103/PhysRevD.94.015019} {\bibfield  {journal} {\bibinfo
  {journal} {Phys. Rev. D}\ }\textbf {\bibinfo {volume} {94}},\ \bibinfo
  {pages} {015019} (\bibinfo {year} {2016})}\BibitemShut {NoStop}%
\bibitem [{\citenamefont {Hochberg}\ \emph {et~al.}(2017)\citenamefont
  {Hochberg}, \citenamefont {Lin},\ and\ \citenamefont {Zurek}}]{Bosonic_4}%
  \BibitemOpen
  \bibfield  {author} {\bibinfo {author} {\bibfnamefont {Y.}~\bibnamefont
  {Hochberg}}, \bibinfo {author} {\bibfnamefont {T.}~\bibnamefont {Lin}}, \
  and\ \bibinfo {author} {\bibfnamefont {K.~M.}\ \bibnamefont {Zurek}},\ }\href
  {\doibase 10.1103/PhysRevD.95.023013} {\bibfield  {journal} {\bibinfo
  {journal} {Phys. Rev. D}\ }\textbf {\bibinfo {volume} {95}},\ \bibinfo
  {pages} {023013} (\bibinfo {year} {2017})}\BibitemShut {NoStop}%
\bibitem [{\citenamefont {Bloch}\ \emph {et~al.}(2017)\citenamefont {Bloch}
  \emph {et~al.}}]{Bosonic_5}%
  \BibitemOpen
  \bibfield  {author} {\bibinfo {author} {\bibfnamefont {I.~M.}\ \bibnamefont
  {Bloch}} \emph {et~al.},\ }\href {\doibase 10.1007/JHEP06(2017)087}
  {\bibfield  {journal} {\bibinfo  {journal} {J. High Energy Phys.}\ }\textbf
  {\bibinfo {volume} {06}},\ \bibinfo {pages} {087} (\bibinfo {year}
  {2017})}\BibitemShut {NoStop}%
\bibitem [{\citenamefont {Green}\ and\ \citenamefont
  {Rajendran}(2017)}]{Bosonic_6}%
  \BibitemOpen
  \bibfield  {author} {\bibinfo {author} {\bibfnamefont {D.}~\bibnamefont
  {Green}}\ and\ \bibinfo {author} {\bibfnamefont {S.}~\bibnamefont
  {Rajendran}},\ }\href {\doibase 10.1007/JHEP10(2017)013} {\bibfield
  {journal} {\bibinfo  {journal} {J. High Energy Phys.}\ }\textbf {\bibinfo
  {volume} {10}},\ \bibinfo {pages} {13} (\bibinfo {year} {2017})}\BibitemShut
  {NoStop}%
\bibitem [{\citenamefont {Arvanitaki}\ \emph {et~al.}(2018)\citenamefont
  {Arvanitaki}, \citenamefont {Dimopoulos},\ and\ \citenamefont
  {Van~Tilburg}}]{Bosonic_7}%
  \BibitemOpen
  \bibfield  {author} {\bibinfo {author} {\bibfnamefont {A.}~\bibnamefont
  {Arvanitaki}}, \bibinfo {author} {\bibfnamefont {S.}~\bibnamefont
  {Dimopoulos}}, \ and\ \bibinfo {author} {\bibfnamefont {K.}~\bibnamefont
  {Van~Tilburg}},\ }\href {\doibase 10.1103/PhysRevX.8.041001} {\bibfield
  {journal} {\bibinfo  {journal} {Phys. Rev. X}\ }\textbf {\bibinfo {volume}
  {8}},\ \bibinfo {pages} {041001} (\bibinfo {year} {2018})}\BibitemShut
  {NoStop}%
\bibitem [{\citenamefont {von Krosigk}\ \emph {et~al.}(2021)\citenamefont {von
  Krosigk} \emph {et~al.}}]{Bosonic_8}%
  \BibitemOpen
  \bibfield  {author} {\bibinfo {author} {\bibfnamefont {B.}~\bibnamefont {von
  Krosigk}} \emph {et~al.},\ }\href {\doibase 10.1103/PhysRevD.104.063002}
  {\bibfield  {journal} {\bibinfo  {journal} {Phys. Rev. D}\ }\textbf {\bibinfo
  {volume} {104}},\ \bibinfo {pages} {063002} (\bibinfo {year}
  {2021})}\BibitemShut {NoStop}%
\bibitem [{\citenamefont {Mitridate}\ \emph {et~al.}(2021)\citenamefont
  {Mitridate}, \citenamefont {Trickle}, \citenamefont {Zhang},\ and\
  \citenamefont {Zurek}}]{Bosonic_9}%
  \BibitemOpen
  \bibfield  {author} {\bibinfo {author} {\bibfnamefont {A.}~\bibnamefont
  {Mitridate}}, \bibinfo {author} {\bibfnamefont {T.}~\bibnamefont {Trickle}},
  \bibinfo {author} {\bibfnamefont {Z.}~\bibnamefont {Zhang}}, \ and\ \bibinfo
  {author} {\bibfnamefont {K.~M.}\ \bibnamefont {Zurek}},\ }\href {\doibase
  10.1007/JHEP09(2021)123} {\bibfield  {journal} {\bibinfo  {journal} {J. High
  Energy Phys.}\ }\textbf {\bibinfo {volume} {09}},\ \bibinfo {pages} {123}
  (\bibinfo {year} {2021})}\BibitemShut {NoStop}%
\bibitem [{\citenamefont {Hochberg}\ \emph {et~al.}(2022)\citenamefont
  {Hochberg} \emph {et~al.}}]{Bosonic_10}%
  \BibitemOpen
  \bibfield  {author} {\bibinfo {author} {\bibfnamefont {Y.}~\bibnamefont
  {Hochberg}} \emph {et~al.},\ }\href {\doibase 10.1103/PhysRevLett.128.191801}
  {\bibfield  {journal} {\bibinfo  {journal} {Phys. Rev. Lett.}\ }\textbf
  {\bibinfo {volume} {128}},\ \bibinfo {pages} {191801} (\bibinfo {year}
  {2022})}\BibitemShut {NoStop}%
\bibitem [{\citenamefont {Dror}\ \emph {et~al.}(2021)\citenamefont {Dror},
  \citenamefont {Elor}, \citenamefont {McGehee},\ and\ \citenamefont
  {Yu}}]{FDEA_2_Dror}%
  \BibitemOpen
  \bibfield  {author} {\bibinfo {author} {\bibfnamefont {J.~A.}\ \bibnamefont
  {Dror}}, \bibinfo {author} {\bibfnamefont {G.}~\bibnamefont {Elor}}, \bibinfo
  {author} {\bibfnamefont {R.}~\bibnamefont {McGehee}}, \ and\ \bibinfo
  {author} {\bibfnamefont {T.~T.}\ \bibnamefont {Yu}},\ }\href {\doibase
  10.1103/PhysRevD.103.035001} {\bibfield  {journal} {\bibinfo  {journal}
  {Phys. Rev. D}\ }\textbf {\bibinfo {volume} {103}},\ \bibinfo {pages}
  {035001} (\bibinfo {year} {2021})}\BibitemShut {NoStop}%
\bibitem [{\citenamefont {Ge}\ \emph {et~al.}(2022)\citenamefont {Ge},
  \citenamefont {He}, \citenamefont {Ma},\ and\ \citenamefont
  {Sheng}}]{FDEA_1_GSF}%
  \BibitemOpen
  \bibfield  {author} {\bibinfo {author} {\bibfnamefont {S.~F.}\ \bibnamefont
  {Ge}}, \bibinfo {author} {\bibfnamefont {X.~G.}\ \bibnamefont {He}}, \bibinfo
  {author} {\bibfnamefont {X.~D.}\ \bibnamefont {Ma}}, \ and\ \bibinfo {author}
  {\bibfnamefont {J.}~\bibnamefont {Sheng}},\ }\href {\doibase
  10.1007/JHEP05(2022)191} {\bibfield  {journal} {\bibinfo  {journal} {J. High
  Energy Phys.}\ }\textbf {\bibinfo {volume} {05}},\ \bibinfo {pages} {191}
  (\bibinfo {year} {2022})}\BibitemShut {NoStop}%
\bibitem [{\citenamefont {Zhang}\ \emph
  {et~al.}(2022{\natexlab{b}})\citenamefont {Zhang} \emph
  {et~al.}}]{FDEA_3_PandaX}%
  \BibitemOpen
  \bibfield  {author} {\bibinfo {author} {\bibfnamefont {D.}~\bibnamefont
  {Zhang}} \emph {et~al.} (\bibinfo {collaboration} {PandaX Collaboration}),\
  }\href {\doibase 10.1103/PhysRevLett.129.161804} {\bibfield  {journal}
  {\bibinfo  {journal} {Phys. Rev. Lett.}\ }\textbf {\bibinfo {volume} {129}},\
  \bibinfo {pages} {161804} (\bibinfo {year} {2022}{\natexlab{b}})}\BibitemShut
  {NoStop}%
\bibitem [{\citenamefont {Cheng}\ \emph {et~al.}(2017)\citenamefont {Cheng}
  \emph {et~al.}}]{CDEX_CJP}%
  \BibitemOpen
  \bibfield  {author} {\bibinfo {author} {\bibfnamefont {J.~P.}\ \bibnamefont
  {Cheng}} \emph {et~al.},\ }\href {\doibase
  10.1146/annurev-nucl-102115-044842} {\bibfield  {journal} {\bibinfo
  {journal} {Annu. Rev. Nucl. Part. Sci.}\ }\textbf {\bibinfo {volume} {67}},\
  \bibinfo {pages} {231} (\bibinfo {year} {2017})}\BibitemShut {NoStop}%
\bibitem [{\citenamefont {She}\ \emph {et~al.}(2020)\citenamefont {She} \emph
  {et~al.}}]{CDEX_SZ}%
  \BibitemOpen
  \bibfield  {author} {\bibinfo {author} {\bibfnamefont {Z.}~\bibnamefont
  {She}} \emph {et~al.} (\bibinfo {collaboration} {CDEX Collaboration}),\
  }\href {\doibase 10.1103/PhysRevLett.124.111301} {\bibfield  {journal}
  {\bibinfo  {journal} {Phys. Rev. Lett.}\ }\textbf {\bibinfo {volume} {124}},\
  \bibinfo {pages} {111301} (\bibinfo {year} {2020})}\BibitemShut {NoStop}%
\bibitem [{\citenamefont {Yang}\ \emph
  {et~al.}(2018{\natexlab{b}})\citenamefont {Yang} \emph
  {et~al.}}]{CDEX_BS_YLT}%
  \BibitemOpen
  \bibfield  {author} {\bibinfo {author} {\bibfnamefont {L.~T.}\ \bibnamefont
  {Yang}} \emph {et~al.},\ }\href {\doibase
  https://doi.org/10.1016/j.nima.2017.12.078} {\bibfield  {journal} {\bibinfo
  {journal} {Nucl. Instrum. Methods Phys. Res., Sect A}\ }\textbf {\bibinfo
  {volume} {886}},\ \bibinfo {pages} {13} (\bibinfo {year}
  {2018}{\natexlab{b}})}\BibitemShut {NoStop}%
\bibitem [{\citenamefont {Baxter}\ \emph {et~al.}(2021)\citenamefont {Baxter}
  \emph {et~al.}}]{DM_rho_1}%
  \BibitemOpen
  \bibfield  {author} {\bibinfo {author} {\bibfnamefont {D.}~\bibnamefont
  {Baxter}} \emph {et~al.},\ }\href {\doibase 10.1140/epjc/s10052-021-09655-y}
  {\bibfield  {journal} {\bibinfo  {journal} {Eur. Phys. J. C}\ }\textbf
  {\bibinfo {volume} {81}},\ \bibinfo {pages} {907} (\bibinfo {year}
  {2021})}\BibitemShut {NoStop}%
\bibitem [{\citenamefont {{J. D. Lewin}}\ and\ \citenamefont {{P. F.
  Smith}}(1996)}]{DM_rho_2}%
  \BibitemOpen
  \bibfield  {author} {\bibinfo {author} {\bibnamefont {{J. D. Lewin}}}\ and\
  \bibinfo {author} {\bibnamefont {{P. F. Smith}}},\ }\href {\doibase
  10.1016/S0927-6505(96)00047-3} {\bibfield  {journal} {\bibinfo  {journal}
  {Astropart. Phys.}\ }\textbf {\bibinfo {volume} {6}},\ \bibinfo {pages} {87}
  (\bibinfo {year} {1996})}\BibitemShut {NoStop}%
\bibitem [{\citenamefont {Catena}\ \emph {et~al.}(2020)\citenamefont {Catena},
  \citenamefont {Emken}, \citenamefont {Spaldin},\ and\ \citenamefont
  {Tarantino}}]{fion_1}%
  \BibitemOpen
  \bibfield  {author} {\bibinfo {author} {\bibfnamefont {R.}~\bibnamefont
  {Catena}}, \bibinfo {author} {\bibfnamefont {T.}~\bibnamefont {Emken}},
  \bibinfo {author} {\bibfnamefont {N.~A.}\ \bibnamefont {Spaldin}}, \ and\
  \bibinfo {author} {\bibfnamefont {W.}~\bibnamefont {Tarantino}},\ }\href
  {\doibase 10.1103/PhysRevResearch.2.033195} {\bibfield  {journal} {\bibinfo
  {journal} {Phys. Rev. Res.}\ }\textbf {\bibinfo {volume} {2}},\ \bibinfo
  {pages} {033195} (\bibinfo {year} {2020})}\BibitemShut {NoStop}%
\bibitem [{\citenamefont {Emken}(2021)}]{DarkART_1}%
  \BibitemOpen
  \bibfield  {author} {\bibinfo {author} {\bibfnamefont {T.}~\bibnamefont
  {Emken}},\ }\href {\doibase DOI:10.5281/zenodo.6046225} {\enquote {\bibinfo
  {title} {{Dark Atomic Response Tabulator (DarkART)[Code, v0.1.0]}},}\
  }\bibinfo {howpublished} {The code can be found under
  \url{https://github.com/temken/darkart}.} (\bibinfo {year}
  {2021})\BibitemShut {NoStop}%
\bibitem [{\citenamefont {Griffin}\ \emph {et~al.}(2021)\citenamefont
  {Griffin}, \citenamefont {Inzani}, \citenamefont {Trickle}, \citenamefont
  {Zhang},\ and\ \citenamefont {Zurek}}]{crystal_1_Griffin}%
  \BibitemOpen
  \bibfield  {author} {\bibinfo {author} {\bibfnamefont {S.~M.}\ \bibnamefont
  {Griffin}}, \bibinfo {author} {\bibfnamefont {K.}~\bibnamefont {Inzani}},
  \bibinfo {author} {\bibfnamefont {T.}~\bibnamefont {Trickle}}, \bibinfo
  {author} {\bibfnamefont {Z.}~\bibnamefont {Zhang}}, \ and\ \bibinfo {author}
  {\bibfnamefont {K.~M.}\ \bibnamefont {Zurek}},\ }\href {\doibase
  10.1103/PhysRevD.104.095015} {\bibfield  {journal} {\bibinfo  {journal}
  {Phys. Rev. D}\ }\textbf {\bibinfo {volume} {104}},\ \bibinfo {pages}
  {095015} (\bibinfo {year} {2021})}\BibitemShut {NoStop}%
\bibitem [{\citenamefont {{C. F. Bunge}}\ \emph {et~al.}(1993)\citenamefont
  {{C. F. Bunge}}, \citenamefont {{J. A. Barrientos}},\ and\ \citenamefont {{A.
  V. Bunge}}}]{bind_eng_Ge}%
  \BibitemOpen
  \bibfield  {author} {\bibinfo {author} {\bibnamefont {{C. F. Bunge}}},
  \bibinfo {author} {\bibnamefont {{J. A. Barrientos}}}, \ and\ \bibinfo
  {author} {\bibnamefont {{A. V. Bunge}}},\ }\href {\doibase
  10.1006/adnd.1993.1003} {\bibfield  {journal} {\bibinfo  {journal} {At. Data
  Nucl. Data Tables}\ }\textbf {\bibinfo {volume} {53}},\ \bibinfo {pages}
  {113} (\bibinfo {year} {1993})}\BibitemShut {NoStop}%
\bibitem [{\citenamefont {Cowan}(1998)}]{chisquare}%
  \BibitemOpen
  \bibfield  {author} {\bibinfo {author} {\bibfnamefont {G.}~\bibnamefont
  {Cowan}},\ }\href {\doibase 10.1093/oso/9780198501565.001.0001} {\emph
  {\bibinfo {title} {Statistical Data Analysis}}}\ (\bibinfo  {publisher}
  {Oxford University Press},\ \bibinfo {address} {New York},\ \bibinfo {year}
  {1998})\BibitemShut {NoStop}%
\bibitem [{\citenamefont {Feldman}\ and\ \citenamefont
  {Cousins}(1998)}]{FC_Method}%
  \BibitemOpen
  \bibfield  {author} {\bibinfo {author} {\bibfnamefont {G.~J.}\ \bibnamefont
  {Feldman}}\ and\ \bibinfo {author} {\bibfnamefont {R.~D.}\ \bibnamefont
  {Cousins}},\ }\href {\doibase 10.1103/PhysRevD.57.3873} {\bibfield  {journal}
  {\bibinfo  {journal} {Phys. Rev. D}\ }\textbf {\bibinfo {volume} {57}},\
  \bibinfo {pages} {3873} (\bibinfo {year} {1998})}\BibitemShut {NoStop}%
\bibitem [{\citenamefont {{Aghanim}}\ \emph {et~al.}(2020)\citenamefont
  {{Aghanim}} \emph {et~al.}}]{CMB_Plank}%
  \BibitemOpen
  \bibfield  {author} {\bibinfo {author} {\bibfnamefont {N.}~\bibnamefont
  {{Aghanim}}} \emph {et~al.} (\bibinfo {collaboration} {Planck
  Collaboration}),\ }\href {\doibase 10.1051/0004-6361/201833910} {\bibfield
  {journal} {\bibinfo  {journal} {Astron. Astrophys.}\ }\textbf {\bibinfo
  {volume} {641}},\ \bibinfo {pages} {A6} (\bibinfo {year} {2020})}\BibitemShut
  {NoStop}%
\end{thebibliography}%

\end{document}